\documentclass[twocolumn,tighten,times]{aastex631}

\usepackage{microtype}
\usepackage{amsfonts}
\usepackage{mathrsfs}
\usepackage{bm}
\usepackage{lipsum}
\usepackage{physics}
\usepackage{multirow}
\usepackage{rotating}
\usepackage{graphicx} 
\usepackage{soul}
\usepackage{subfigure}
\usepackage{color}

\shorttitle{Classification of KiDS: QSO-Star-Galaxy}
\shortauthors{Feng et al.}

\begin{document}

\title{\bf \large Morpho-Photometric Classification of KiDS DR5 Sources Based on Neural Networks: A Comprehensive Star-Quasar-Galaxy Catalog}

\correspondingauthor{Hai-Cheng Feng, Rui Li, \& Nicola R. Napolitano}
\email{hcfeng@ynao.ac.cn} \email{liruiww@gmail.com} \email{nicolarosario.napolitano@unina.it}

\author[0000-0002-1530-2680]{Hai-Cheng Feng}
\affiliation{Yunnan Observatories, Chinese Academy of Sciences, Kunming 650216, Yunnan, People's Republic of China}
\affiliation{Key Laboratory for the Structure and Evolution of Celestial Objects, Chinese Academy of Sciences, Kunming 650216, Yunnan, People's Republic of China}
\affiliation{Center for Astronomical Mega-Science, Chinese Academy of Sciences, 20A Datun Road, Chaoyang District, Beijing 100012, People's Republic of China}
\affiliation{Key Laboratory of Radio Astronomy and Technology, Chinese Academy of Sciences, 20A Datun Road, Chaoyang District, Beijing 100101, People's Republic of China}

\author[0000-0002-3490-4089]{Rui Li}
\affiliation{Institude for Astrophysics, School of Physics, Zhengzhou University, Zhengzhou, 450001, People's Republic of China}

\author[0000-0003-0911-8884]{Nicola R. Napolitano}
\affiliation{School of Physics and Astronomy, Sun Yat-sen University, Zhuhai Campus, 2 Daxue Road, Xiangzhou District, Zhuhai, People's Republic of China}
\affiliation{CSST Science Center for Guangdong-Hong Kong-Macau Great Bay Area, Zhuhai, 519082, People's Republic of China}
\affiliation{INAF -- Osservatorio Astronomico di Capodimonte, Salita Moiariello 16, 80131 - Napoli, Italy}

\author[0000-0003-3823-3419]{Sha-Sha Li}
\affiliation{Yunnan Observatories, Chinese Academy of Sciences, Kunming 650216, Yunnan, People's Republic of China}
\affiliation{Key Laboratory for the Structure and Evolution of Celestial Objects, Chinese Academy of Sciences, Kunming 650216, Yunnan, People's Republic of China}
\affiliation{Center for Astronomical Mega-Science, Chinese Academy of Sciences, 20A Datun Road, Chaoyang District, Beijing 100012, People's Republic of China}
\affiliation{Key Laboratory of Radio Astronomy and Technology, Chinese Academy of Sciences, 20A Datun Road, Chaoyang District, Beijing 100101, People's Republic of China}
\author{J. M. Bai}
\affiliation{Yunnan Observatories, Chinese Academy of Sciences, Kunming 650216, Yunnan, People's Republic of China}
\affiliation{Key Laboratory for the Structure and Evolution of Celestial Objects, Chinese Academy of Sciences, Kunming 650216, Yunnan, People's Republic of China}
\affiliation{Center for Astronomical Mega-Science, Chinese Academy of Sciences, 20A Datun Road, Chaoyang District, Beijing 100012, People's Republic of China}
\affiliation{Key Laboratory of Radio Astronomy and Technology, Chinese Academy of Sciences, 20A Datun Road, Chaoyang District, Beijing 100101, People's Republic of China}
\author[0009-0002-1296-4256]{Yue Dong}
\affiliation{School of Mathematics and Physics, Xi’an Jiaotong-Liverpool University, 111 Renai Road, Suzhou, 215123, PR China}
\author{Ran Li}
\affiliation{University of Chinese Academy of Sciences, Beijing 100049, People's Republic of China}
\affiliation{National Astronomical Observatories, Chinese Academy of Sciences, 20A Datun Road, Chaoyang District, Beijing 100012, People's Republic of China}

\author[0000-0002-2153-3688]{H. T. Liu}
\affiliation{Yunnan Observatories, Chinese Academy of Sciences, Kunming 650216, Yunnan, People's Republic of China}
\affiliation{Key Laboratory for the Structure and Evolution of Celestial Objects, Chinese Academy of Sciences, Kunming 650216, Yunnan, People's Republic of China}
\affiliation{Center for Astronomical Mega-Science, Chinese Academy of Sciences, 20A Datun Road, Chaoyang District, Beijing 100012, People's Republic of China}

\author[0000-0002-2310-0982]{Kai-Xing Lu}
\affiliation{Yunnan Observatories, Chinese Academy of Sciences, Kunming 650216, Yunnan, People's Republic of China}
\affiliation{Key Laboratory for the Structure and Evolution of Celestial Objects, Chinese Academy of Sciences, Kunming 650216, Yunnan, People's Republic of China}
\affiliation{Center for Astronomical Mega-Science, Chinese Academy of Sciences, 20A Datun Road, Chaoyang District, Beijing 100012, People's Republic of China}

\author[0000-0003-0230-6436]{Zhi-Wei Pan}
\affiliation{Department of Astronomy, School of Physics, Peking University, Beijing 100871, People's Republic of China}
\affiliation{Kavli Institute for Astronomy and Astrophysics, Peking University, Beijing 100871, People's Republic of China}

\author{Mario Radovich}
\affiliation{INAF - Osservatorio Astronomico di Padova, via dell'Osservatorio 5, 35122 Padova, Italy}

\author{Huan-Yuan Shan}
\affiliation{Shanghai Astronomical Observatory, Chinese Academy of Sciences, Shanghai 200030, People's Republic of China}

\author[0000-0003-4156-3793]{Jian-Guo Wang}
\affiliation{Yunnan Observatories, Chinese Academy of Sciences, Kunming 650216, Yunnan, People's Republic of China}
\affiliation{Key Laboratory for the Structure and Evolution of Celestial Objects, Chinese Academy of Sciences, Kunming 650216, Yunnan, People's Republic of China}
\affiliation{Center for Astronomical Mega-Science, Chinese Academy of Sciences, 20A Datun Road, Chaoyang District, Beijing 100012, People's Republic of China}

\author{Wen-Zhe Xi}
\affiliation{Yunnan Observatories, Chinese Academy of Sciences, Kunming 650216, Yunnan, People's Republic of China}
\affiliation{University of Chinese Academy of Sciences, Beijing 100049, People's Republic of China}

\author{Ling-Hua Xie}
\affiliation{School of Physics and Astronomy, Sun Yat-sen University, Zhuhai Campus, 2 Daxue Road, Xiangzhou District, Zhuhai, People's Republic of China}

\author{Zun-Li Yuan}
\affiliation{Department of Physics, School of Physics and Electronics, Hunan Normal University, Changsha 410081, China}

\author{Yang-Wei Zhang}
\affiliation{South-Western Institute for Astronomy Research, Yunnan University, Kunming 650500, People's Republic of China}

\begin{abstract}

We present a novel multimodal neural network (MNN) for classifying astronomical sources in multiband ground-based observations, from optical to near infrared, to separate sources in stars, galaxies and quasars. Our approach combines a convolutional neural network branch for learning morphological features from $r$-band images with an artificial neural network branch for extracting spectral energy distribution (SED) information. Specifically, we have used 9-band optical ($ugri$) and NIR ($ZYHJK_s$) data from the Kilo-Degree Survey (KiDS) Data Release 5. The two branches of the network are concatenated and feed into fully-connected layers for final classification. We train the network on a spectroscopically confirmed sample from the Sloan Digital Sky Survey cross-matched with KiDS. The trained model achieves 98.76\% overall accuracy on an independent testing dataset, with F1 scores exceeding 95\% for each class. Raising the output probability threshold, we obtain higher purity at the cost of a lower completeness. We have also validated the network using external catalogs cross-matched with KiDS, correctly classifying 99.74\% of a pure star sample selected from Gaia parallaxes and proper motions, and 99.74\% of an external galaxy sample from the Galaxy and Mass Assembly survey, adjusted for low-redshift contamination. We apply the trained network to 27,335,836 KiDS DR5 sources with $r \leqslant 23$ mag to generate a new classification catalog. This MNN successfully leverages both morphological and SED information to enable efficient and robust classification of stars, quasars, and galaxies in large photometric surveys.

\end{abstract}



\section{Introduction} \label{sec:1}

Stars, galaxies, and quasars are the three main categories of astronomical objects in most large scale surveys, enabling advances in our understanding of cosmic structure, evolution, and fundamental physics. Stars, the most observable entities in the universe, are the basic building blocks of galaxies. Systematic observations and cataloging of stars not only yield a wealth of information about stellar evolution and the synthesis of heavy elements, but also aid in mapping the structure and dynamics of the Milky Way \citep{Helmi2020, Zhang2023}. Galaxies, large assemblies of stars, gas, dust, and dark matter bound by gravity, serve as fundamental components of large-scale structure. A comprehensive census and analysis of galaxies can reveal the underlying distribution of dark matter and provide a platform for testing cosmological models that describe the origin and evolution of the Universe \citep{Bond1996, Behroozi2013}. Quasars, also known as quasi-stellar objects (QSOs), represent extremely luminous galactic nuclei fueled by matter accretion onto supermassive black holes. They serve as invaluable tools for probing the formation and evolution of supermassive black holes, and understanding their connection to their host galaxies \citep{Schmidt1963, Salpeter1964, Kormendy2013}. Accurate classification of these diverse celestial objects stands as an indispensable prerequisite for conducting detailed follow-up studies and science analyses.

Over the course of several decades, extensive efforts have been devoted to develop various approaches for separating stars, QSOs, and galaxies. Spectroscopic observations are among the most reliable methods, primarily due to the unique physical properties of these three types of astronomical objects \citep[e.g.,][]{Baldwin1981, Feng2021, Verro2022}. The emergence of spectroscopic survey projects, such as the Sloan Digital Sky Survey \citep[SDSS;][]{Yo00}, the Large Sky Area Multi-Object Fiber Spectroscopic Telescope \citep[LAMOST;][]{Cui2012}, and the Dark Energy Spectroscopic Instrument \citep[DESI;][]{DESICollaboration2016}, has enabled the collection of spectra for millions of targets. This has, in turn, led to the proposal of some automatic spectral classification methods, including the Baldwin-Phillips-Terlevich (BPT) diagnostic scheme \citep{Baldwin1981} and template matching \citep{Bolton2012}. However, most spectroscopic surveys employ multi-object fiber spectrographs, which are limited to observing hundreds to thousands of targets concurrently, necessitating exposure times of several thousand seconds. This requires a significant amount of telescope time to obtain large statistical samples, but results in relatively shallow data for each target. As a result, target pre-selection is often implemented in these surveys, which can introduce bias in sample statistics and increase the likelihood of missing peculiar celestial populations.

As alternative to spectroscopy, multi-band imaging can effectively trace the impact of different radiative process in the spectral energy distributions (SEDs) of astronomical sources. In this case, source classification can be performed via color-color diagnostic \citep{Richards2002, Peters2015} or SED fitting \citep{Ilbert2009, Salvato2009}. The photometric approach has the advantage to collect all targets under a limited brightness in the field of view, effectively eliminating any selection bias. In contrast to spectroscopic observation, photometry requires shorter exposure times, often just a few minutes, to reach deeper limiting magnitudes. Furthermore, image data can provide morphological details that are not accessible from spectra, which contains rich physical information. Indeed, the first attempt to perform object classifications, the primary methods were visual inspection of images or analysis of morphological parameters \citep[e.g.,][]{Hubble1926, MacGillivray1976, Yee1991, Scranton2002, Kelvin2014, Lopez-Sanjuan2019}. This technique proved to be effective in differentiating between stars (point sources) and galaxies (extended structures), and even enabled the identification of certain QSOs with notable host galaxies \citep{Guo2022}.

Ongoing wide-field imaging surveys provide continuous streams of multi-band photometric data, enabling the potential for obtaining complete samples of various celestial object. Current projects like the Hyper Suprime-Cam Subaru Strategic Program \citep[HSC-SSP;][]{Aihara2018}, Dark Energy Survey \citep[DES;][]{DESCollaboration2005}, and Kilo Degree Square Survey \citep[KiDS;][]{de15} are amassing observations spanning hundreds of millions of targets. Upcoming programs, like the China Space Station Telescope \citep[CSST;][]{Gong2019}, Large Synoptic Survey Telescope \citep[LSST;][]{Ivezic2019}, and the Euclid mission \citep{Laureijs2011}, will dramatically escalate data volumes into billions or more. The wealth of datasets underscores the urgent need for rapid and efficient classification algorithms.

SED-based classification methods typically require a broad wavelength coverage. Depending on the classes one wants to classify these might include  other bands like X-rays, infrared or radio. However, most survey projects operate within relatively narrow wavelength ranges. For instance, HSC and DES focus on optical bands, the Two Micron All Sky Survey \citep[2MASS;][]{Skrutskie2006} on near-infrared (NIR) bands, and the Wide-Field Infrared Survey Explorer \citep[WISE;][]{Wright2010} on mid-infrared bands. Classifying on the basis of photometric data points with short wavelength intervals may be affected by degeneracy \citep{Buchs2019}. One approach to address this is to combine multiple survey data \citep[e.g.,][]{Ilbert2009, Fotopoulou2016, Khramtsov2021, Salvato2022, Yang2023}, but this approach can only be applied to targets with overlapping regions and is limited to the depth of the shallowest survey. Traditional morphology-based classification demands high-quality data where low-surface brightness galaxies may be missed. Additionally, classifications suffer intrinsic degeneracies like compact galaxies and high-redshift QSOs confused as point sources. Combining the SED of celestial objects with their morphological features holds promise in mitigating degeneracies during the classification process and fully exploring the scientific potential of survey data \citep{Sevilla-Noarbe2018, Khramtsov2019, Nakoneczny2019, Nakoneczny2021}.

Machine learning (ML) technology, with its outstanding speed and accuracy, has been widely applied in the field of astronomical big data processing \citep[e.g.,][]{Dieleman2015, Bai2019, Logan2020, Szklenar2020, Chen2021, Zhou2021, Dubois2022, Li2022, Xie2023}. Unlike traditional methods relying on physical modeling, ML can directly learn and mine the intrinsic features of celestial objects from vast observational data, sometimes even making new discoveries beyond existing theories. Currently, many ML algorithms have been successfully utilized in astronomical classification tasks. Techniques like Random Forests, extreme gradient boosting (XGBoost), support vector machines, and artificial neural networks (ANN) are commonly used for analyzing catalogs of photometric or morphological parameters, while convolutional neural networks (CNN) can directly extract information from images or spectra \citep{Cabayol2019, Li2019, Tadaki2020, Nakazono2021, Chaini2023}.

Although some studies have attempted to feed both SED and morphological information into ML models \citep{Fadely2012, Khramtsov2019, Nakoneczny2021}, most studies still rely on parameterized morphological parameters obtained through traditional methods, which can be affected by random noise. Meanwhile, these methods often face the following limitations: (1) they are generally more suited for small-scale datasets, with performance plateauing as dataset size increases; (2) they struggle to model complex patterns that require deep and hierarchical representations; and (3) they require manual feature engineering, which is labor-intensive and prone to biases.

Motivated by the potential to enhance classification performance, we construct an multimodal neural network (MNN) that combines the flexibility of ANN and CNN to simultaneously extract SED and image features, respectively. We have applied this network to the KiDS DR5 dataset to generate a new Star-QSO-Galaxy catalog. In Section \ref{sec:2}, we provide a brief introduction to the data from each survey. Section \ref{sec:3} presents the methodology, detailing the construction and training of the MNN. The results are presented in Section \ref{sec:4}, followed by a discussion in Section \ref{sec:5} and concluding remarks in Section \ref{sec:6}.

\begin{figure*}[!ht]
\centering
\includegraphics[scale=0.5]{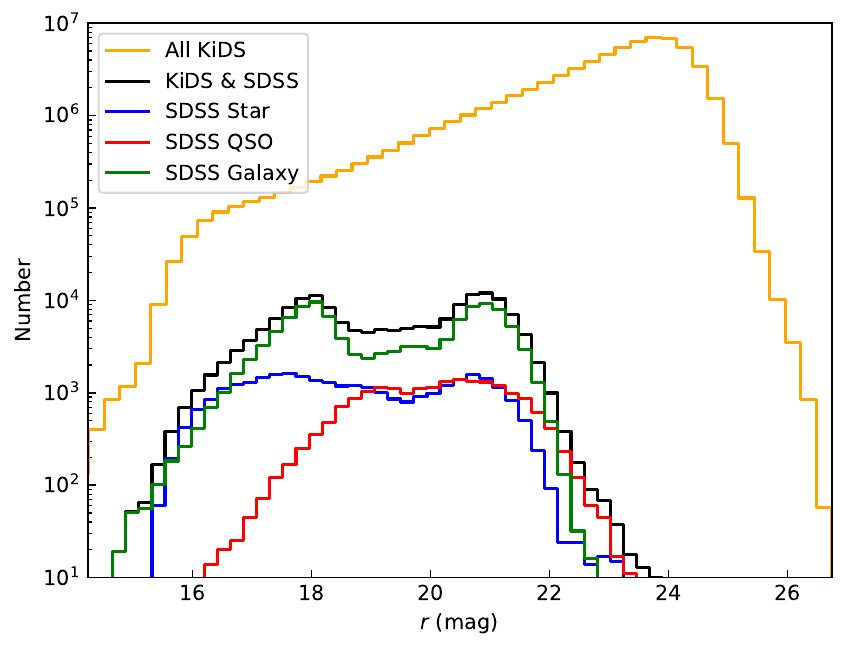}
\includegraphics[scale=0.5]{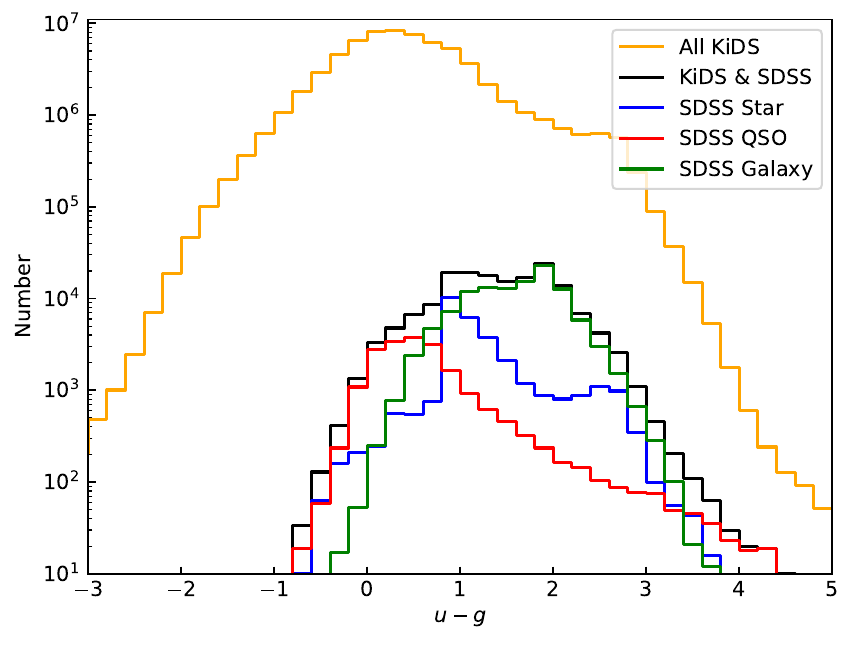}
\caption{The distributions of $r$-band magnitude (left panel) and $u-g$ color (right panel) for KiDS and SDSS datasets. In both panels, the black histograms represent the overlapping sources between KiDS and SDSS, the orange histograms depict the data from the KiDS survey, while the blue, red, and green histograms denote stars, QSOs, and galaxies identified by SDSS spectroscopy, respectively.
} 
\label{fig1}
\end{figure*}

\section{Data} \label{sec:2}

The aim of this paper is to provide a Galaxy-QSO-Star catalog for KiDS public Data Release 5 (DR5). Ground-truth labels required for supervised ML come from SDSS DR17. Additionally, we adopt Gaia \citep{Ga16} DR3 and Galaxy and Mass Assembly \citep[GAMA;][]{Dr11} DR4 to further test the performance of classification algorithm to Galactic (stars) and extragalactic (QSOs and galaxies) objects, respectively.

\subsection{KiDS} \label{sec:2.1}

KiDS is a $ugri$-band wide-field imaging survey with the OmegaCAM camera mounted at the VLT Survey Telescope \citep{Ca11, Ca12}. The complete observations cover $\sim$1350 deg$^{2}$ in the Galactic North and South. 1006 deg$^{2}$ of them have been released under DR4 \citep{Ku19}, while the remaining DR5 is internal data. Thanks to the excellent observing conditions, the $r$-band limiting magnitude is $\sim$25 (5$\sigma$ in 2 square arcsec) with typical seeing of $\sim$0.7\farcs\ In addition, VISTA Kilo-degree Infrared Galaxy \citep[VIKING;][]{Ed13} is KiDS partner survey carried out with the VISTA telescope which can complement its near-infrared data in $ZYHJK_{s}$-band. Therefore, we can obtain a total of 9-band optical-NIR data to separate QSOs from stars and galaxies.

For each object, KiDS DR5 provides its 9-band Gaussian Aperture and PSF \citep[GAaP;][]{Ku15} magnitudes, and then 8 adjacent band colors (e.g., $u - g$, $g - r$, etc.) can be derived. The GAaP technique is specifically designed to improve color measurement accuracy by homogenizing the PSF across different filters. This is especially important for galaxies and some nearby QSOs. To avoid any confusion introduced by the unreliable measurements, we only adopt the 65,909,027 objects which have all 9-band measurements. Figure~\ref{fig1} shows the distribution of their $r$-band GAaP magnitude and corresponding $u - g$ color. Considering that supervised machine learning algorithms are limited by the feature space coverage of training samples (see Figure~\ref{fig1}), we further select the 27,335,836 objects with $r \leqslant$ 23 mag as inference data (see Section~\ref{sec:4}). In principle, the reliability of the magnitude measurement should be characterized by the error or the signal-to-noise ratio (S/N), but this can be ignored for bright sources.

\subsection{SDSS} \label{sec:2.2}

DR17 is the final annual release of SDSS-IV \citep{Ab22} which has cumulated 5,801,200 spectroscopically identified objects in specObj-dr17\footnote{\url{https://data.sdss.org/datamodel/files/SPECTRO_REDUX/specObj.html}} data base. We apply TOPCAT tool \citep{Ta05} to cross-match SDSS sample with KiDS, and 177,843 objects are obtained within a matching radius of 1 arcsec. To ensure the reliability of SDSS identification, we only select the confident spectroscopic classification measurements depending on the ZWARNING flag \citep[ZWARNING = 0;][]{Bolton2012}, which yield 167,009 objects. The final cross-matched sample includes 31,306 stars, 19,608 QSOs, and 116,095 galaxies. We then randomly split a subset of 127,009 (76\%) as the training dataset to train the MNN model described in Section~\ref{sec:3}, while the remaining 40,000 (24\%) sources are equally divided into validation and testing datasets. Note that the inference data are deeper than the labeled sample (see Figure~\ref{fig1}), thus it is necessary to test the performance of our algorithm.

\subsection{Gaia} \label{sec:2.3}

Gaia is an European Space Agency’s all-sky space-astrometry mission launched on 19 December 2013. Its primary science goal is to study the structure and dynamics of the Galaxy through high precision astrometry, which can help us to isolate a clean sample of star with significant parallax or proper motion. There are 1,811,709,771 sources brighter than $G$ = 21 (corresponding to $r \approx$ 20) in the Gaia DR3 catalog \citep{Ga21, Ga22}, and 4,840,178 of them can be matched to KiDS within 1 arcsec through the CDS Upload X-Match Window of TOPCAT. Then, the most confident stars are selected by the significance of parallax and proper motion. We required that a star in the sample should satisfy either of the following criteria \citep[see also in][]{Yang2023}:
\begin{equation}
\frac{|\varpi|}{\sigma_{\varpi}} \geqslant 5, 
\label{Eq1}
\end{equation}
or
\begin{equation}
\frac{\mu^{2}_{\alpha} + \mu^{2}_{\delta}}{\sqrt{\mu^{2}_{\alpha}\sigma^{2}_{\mu_{\alpha}} + \mu^{2}_{\delta}\sigma^{2}_{\mu_{\delta}}}} \geqslant 5, 
\label{Eq2}
\end{equation}
where $\varpi$, $\sigma_{\varpi}$, $\mu_{\alpha}$, $\sigma_{\mu_{\alpha}}$, $\mu_{\delta}$, and $\sigma_{\mu_{\delta}}$ are parallax, parallax error, proper motion in RA, proper motion error in RA, proper motion in DEC, and proper motion error in DEC, respectively. This selection step results in a sample of 3,444,939. The final Gaia star dataset contains 3,422,113 sources after removing 22,826 SDSS objects.

\subsection{GAMA} \label{sec:2.4}

GAMA is a spectroscopic survey carried out on the 3.9-m Anglo-Australian Telescope which is designed to study the formation and evolution of cosmology and galaxy. The final data release of GAMA, DR4 \citep{Dr22}, contains reliable redshift measurements (NQ $>$ 2) for 330,542 sources. Most of them are galaxies, while some stars and QSOs can be distinguished by redshift (e.g., the objects with redshifts $z \sim 0$ and $z > 1$ are most likely to be stars and QSOs, respectively). We cross-matched KiDS with GAMA using a radius of 1 arcsec, yielding a sample of 279,404. However, 48,902 of them are duplicated with SDSS observations which may cause bias in the test results. To maintain an independent GAMA sample, we removed these objects, leaving a total of 230,502 sources.

\section{Method} \label{sec:3}
We employed two standard models, ANN and CNN, which are significant branches of ML \citep{Smith2023}. These models are designed based on the concept of emulating the operations of biological neural networks, which in turn allow for the efficient processing and analysis of complex data. ANN \citep{McCulloch1943}, also referred to as Multi-Layer Perceptron (MLP), comprises multiple interconnected layers of artificial neurons that mimic the connections and synaptic processing of biological brains. When an ANN comprises multiple hidden layers, it belongs to the category of deep neural networks (DNNs). Each neuron learns weighting and bias parameters to capture nonlinear relationships within inputs, thus enabling accurate prediction and classification. CNN \citep{Lecun1998}, on the other hand, is a specialized neural network architecture that is inherently a part of the DNN family. It is particularly effective in image and signal processing tasks. By employing components such as convolutional layers, pooling layers, and fully connected layers, CNN is capable of extracting spatial and morphological features from input data. It accomplishes this through weight sharing and local perception mechanisms. As a result, CNN exhibits superior performance in tasks like image classification and object detection, where its ability to analyze complex visual information is particularly advantageous.

\begin{figure*}[!ht]
\centering
\includegraphics[scale=0.5]{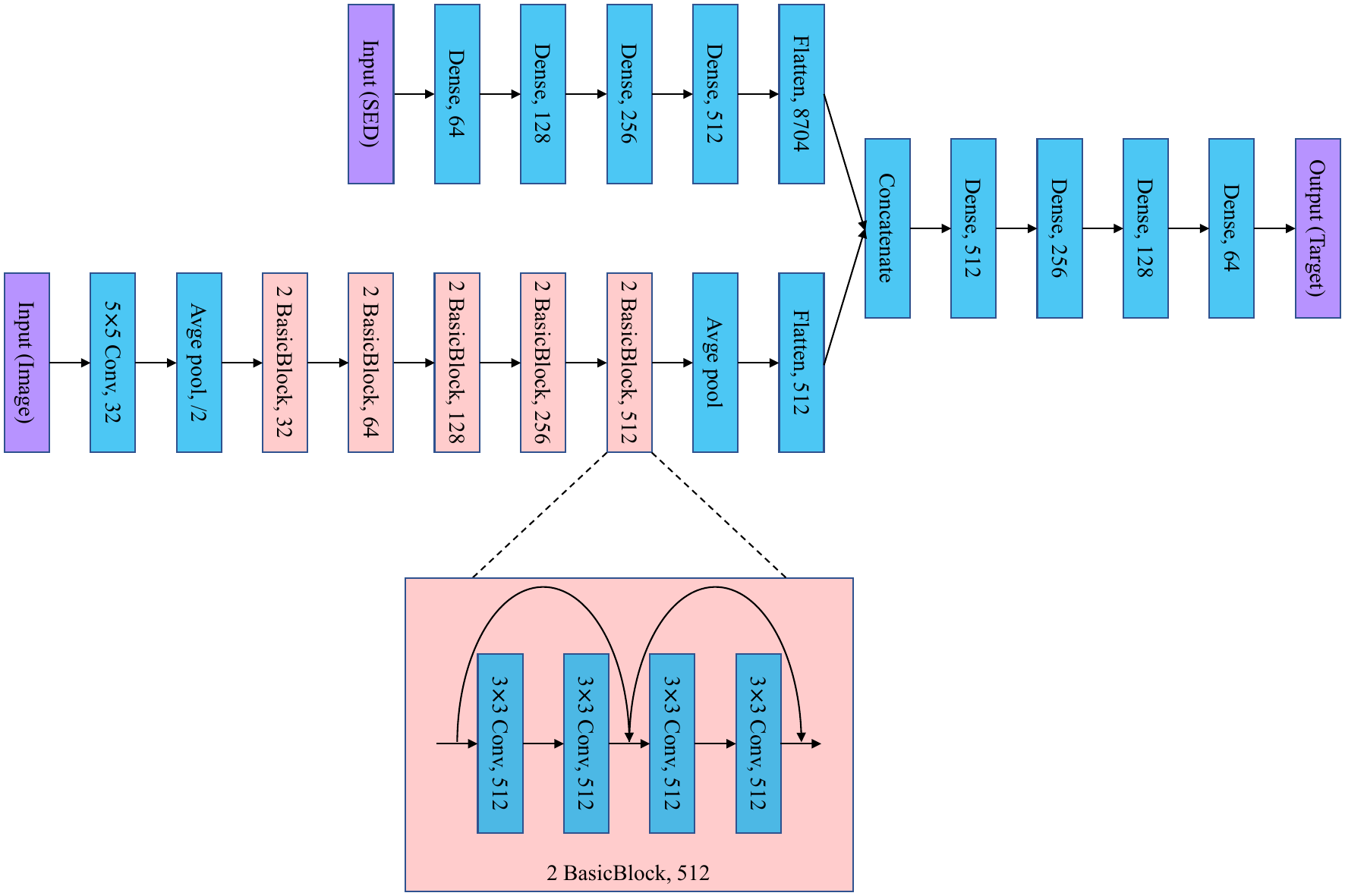}
\caption{Schematic of the MNN architecture, consisting of 18,257,155 neurons. Each blue or purple rectangle represents a layer, and each light pink rectangle represents two identical ResNet blocks. The SED branch inputs photometry from 9 KiDS bands and 8 adjacent band colors, and the image branch processes a $41\times41$ pixels $r$-band image. The output layer provides classification probabilities for stars, quasars, or galaxies.} 
\label{fig2}
\end{figure*}

\subsection{Neural network architecture} \label{sec:3.1}

As illustrated in Figure~\ref{fig2}, our MNN integrates two simultaneous branches tailored for feature extraction from SED and image data. These branches aim to capture different attributes of the data: the ANN is tasked with analyzing complex nonlinear patterns in multi-band photometric information, while the CNN focuses on extracting spatial structures and morphological features from astronomical images. The high-level features extracted from both data streams are combined to form a complementary feature vector that feeds into the final classification layers, thereby improving the overall accuracy of our model.

The ANN branch begins with an input layer, where it receives photometric information spanning nine bands along with eight colors derived from adjacent bands. Following this are four fully connected layers, consisting of 64, 128, 256, and 512 neurons, respectively. These dense layers are dedicated to capturing the complex non-linear relationships embedded within the SED data. The output from the final dense layer is subsequently flattened into a one-dimensional vector containing 8704 elements.

In parallel, the CNN branch operates on a $41\times41$ pixels (8$^\prime$ $\times$ 8$^\prime$) $r$-band image input, initiating the feature extraction with a $5\times5$ convolutional layer comprising 32 channels. This layer is primarily responsible for capturing basic visual features from the image data. This is followed by an average pooling layer, which halves the size of the feature maps while preserving the most critical information. To facilitate deeper learning and more robust feature detection, we incorporate five ResNet \citep{He15} basic blocks in succession. These blocks are arranged to address the vanishing or exploding gradient issues by implementing shortcut connections after every two convolutional layers. Each basic block is comprised of four convolutional layers with $3\times3$ kernels, with the number of filters doubling at each subsequent block—beginning with 32 and incrementing through 64, 128, 256, to 512. The final average pooling layer further compresses the feature dimensions by half, then passing it to a flattening layer that forms another one-dimensional 512-feature vector.

The architecture of the MNN was developed through a systematic and iterative process. For the classification of SEDs, we first evaluated dense networks with 1 to 15 layers and found that performance stabilized beyond 4 layers. Based on this result, a 4-layer dense architecture was adopted to construct the MLP component of the network. This MLP was then concatenated with the ResNet18 backbone  to enable multimodal feature fusion for classification. During this stage, we systematically explored various CNN hyperparameters, such as the number of layers, pooling strategies, and kernel sizes, and found that some adjustments contributed to performance improvements. Additionally, we examined the addition of dense layers at the output stage, which yielded further gains. The final architecture was refined by fine-tuning parameters across the entire multimodal network to achieve optimal results. Throughout the design process, we prioritized computational efficiency by minimizing the number of parameters when performance differences were negligible. The performance differences across most network configurations were typically less than 1\%.

The feature vectors from the ANN and CNN—8704 and 512 dimensions, respectively—are concatenated to form a comprehensive 9216-dimensional vector that incorporates both the photometric and morphological characteristics of the astronomical sources. This integrated vector is then refined through a series of dense layers with a descending number of neurons (512, 256, 128, and 64), effectively reducing the dimensionality of the feature space. The progressive dimensionality reduction may serve multiple purposes within the network. For example, it helps mitigate the risk of overfitting by simplifying the feature space, facilitates the integration and transformation of features from different modalities, and improves computational efficiency by reducing the overall complexity of the network \citep{Bengio2012}. Culminating the network is a final dense layer with three neurons, corresponding to the categories of stars, galaxies, and QSOs. This output layer employs a softmax activation function to produce a probabilistic distribution over the three classes, facilitating the determination of the most likely category for each astronomical source.

Throughout the network, each hidden layer applies the ReLU activation function to introduce non-linearity and facilitate effective gradient propagation, which is vital for maintaining efficient training of deep neural networks \citep{Nair2010, Glorot2011}. We adopted the categorical cross-entropy as our loss function and utilized the Adam optimizer to refine the training of our model.

\begin{figure*}[!ht]
\centering
\includegraphics[scale=0.45]{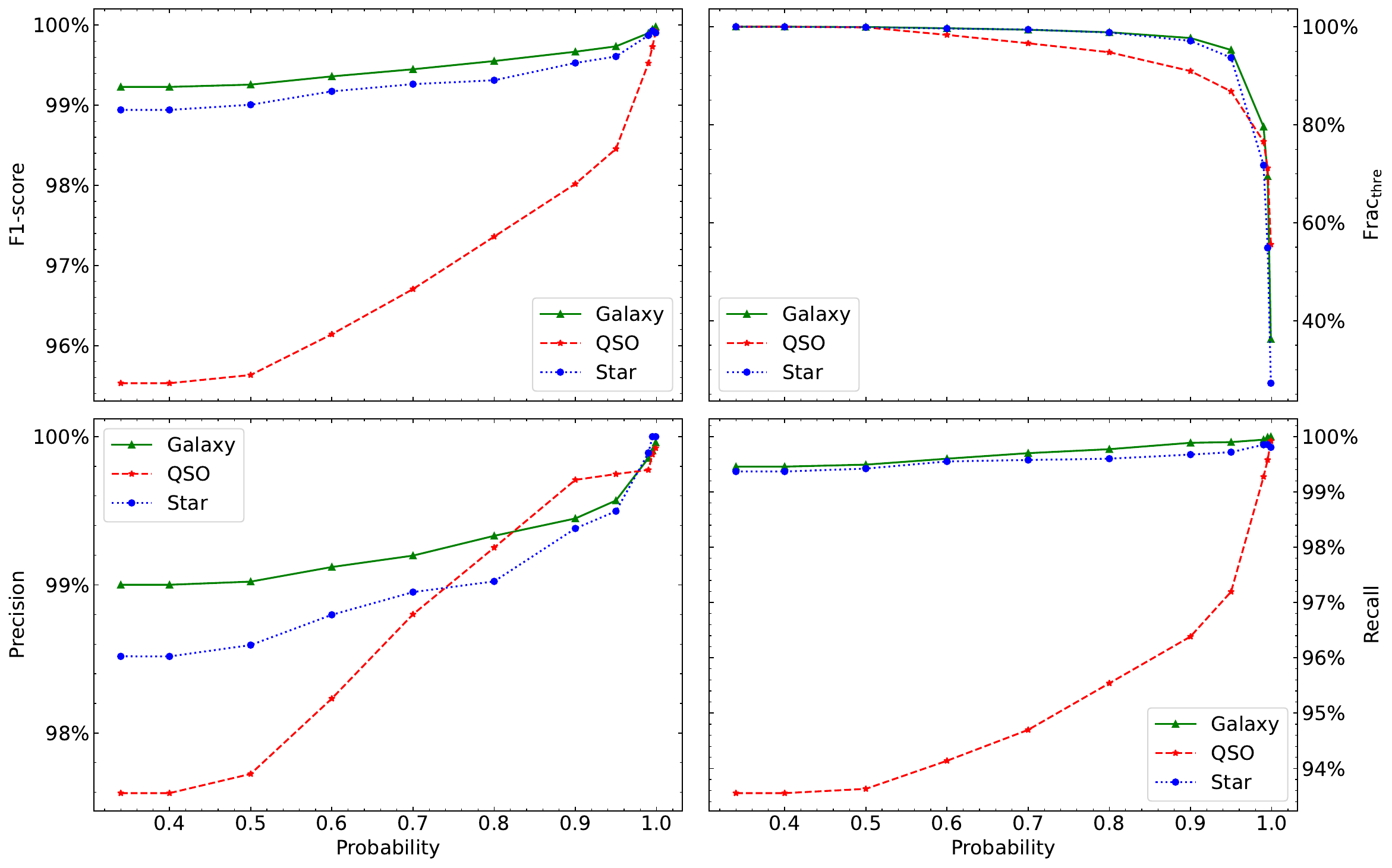}
\caption{Evaluation metrics for classification of stars, quasars, and galaxies at different output probability thresholds. The panels display F1-score (top left), $\rm Frac_{thre}$ (top right), Precision (bottom left), and Recall (bottom right), with blue, red, and green lines representing the metrics for stars, quasars, and galaxies, respectively.} 
\label{fig3}
\end{figure*}

\begin{figure}[!ht]
\centering
\includegraphics[scale=0.55]{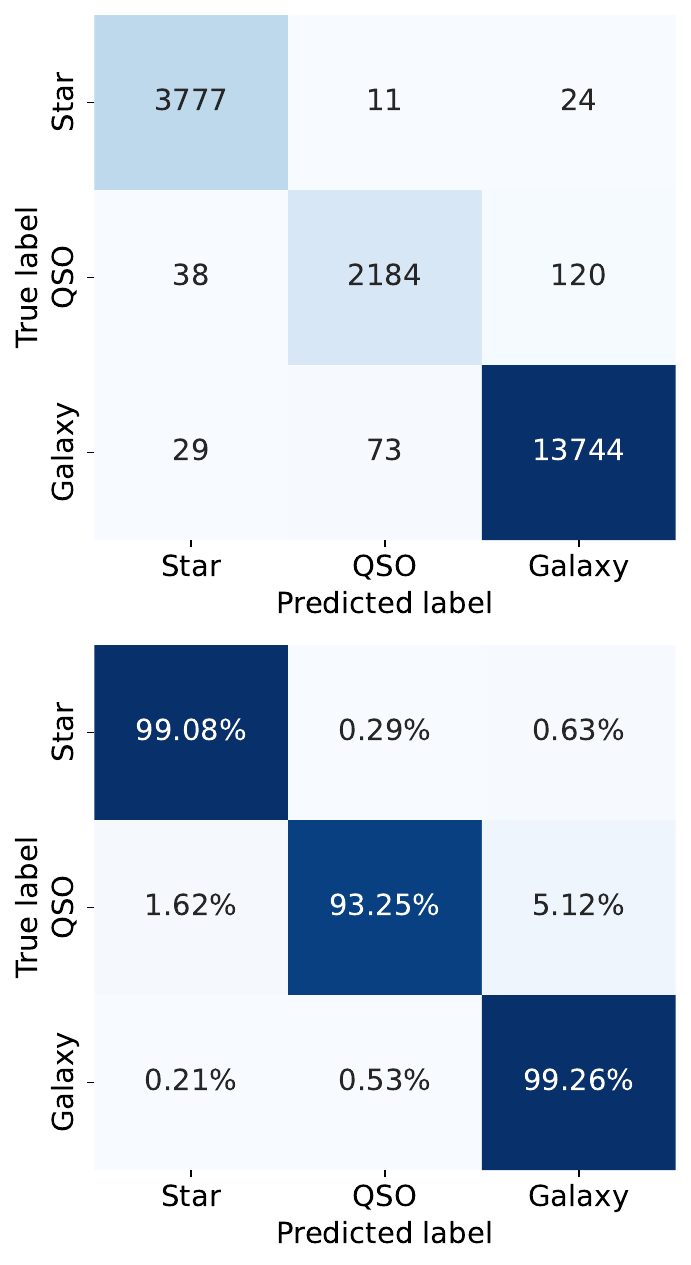}
\caption{Confusion matrix for the internal testing dataset. Each panel has predicted categories (star, quasar, galaxy) on the vertical axis and true categories from SDSS on the horizontal axis. The top panel shows the number of targets predicted in each category, while the bottom panel shows these numbers normalized as percentages.} 
\label{fig4}
\end{figure}

\begin{figure}[!ht]
\centering
\includegraphics[scale=0.55]{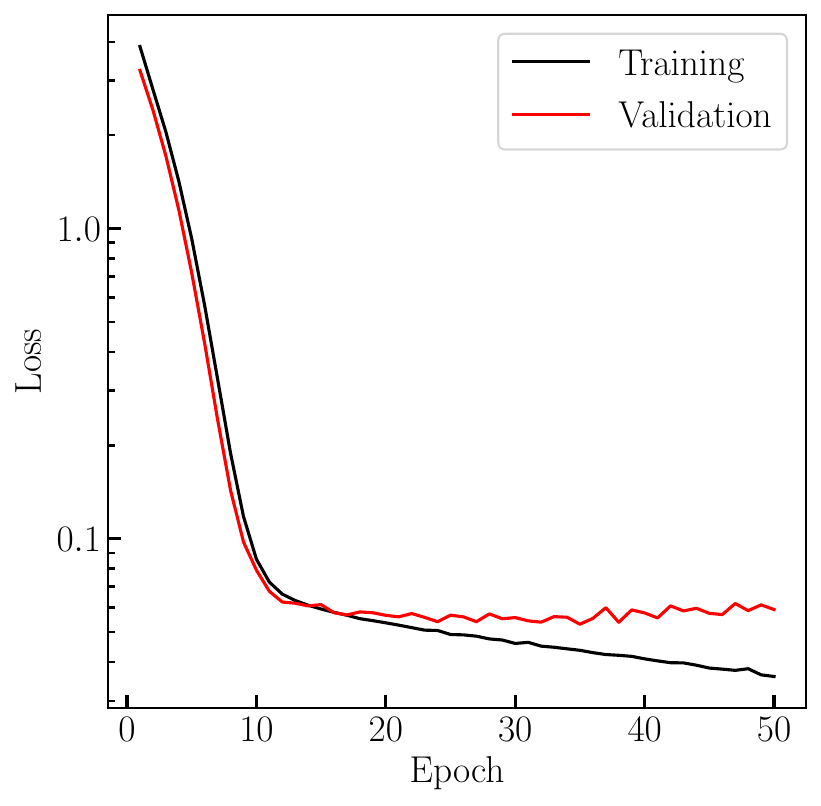}
\includegraphics[scale=0.55]{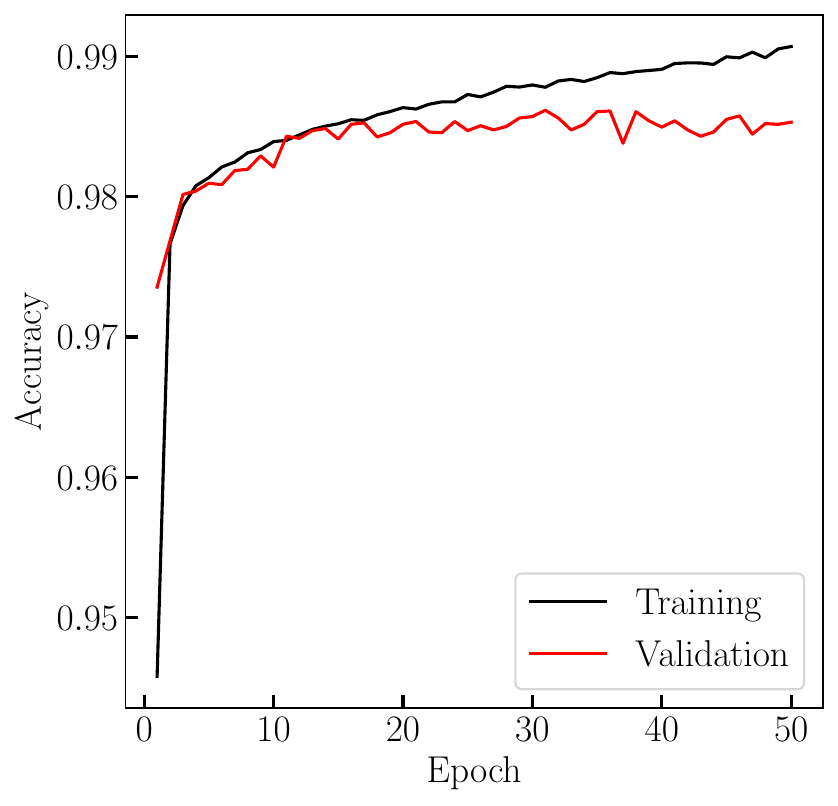}
\caption{Training and validation loss (top) and accuracy (bottom) as a function of epochs.} 
\label{fig5}
\end{figure}

\subsection{Evaluation metrics} \label{sec:3.2}
In ML, several standard metrics are commonly employed to assess classification performance, including Accuracy, Precision, Recall, and F1-score. These metrics are defined in the following manner:
\begin{equation}
{\rm Accuracy = \frac{TP + TN}{TP + TN + FP + FN}}, 
\label{Eq3}
\end{equation}

\begin{equation}
{\rm Precision = \frac{TP}{TP + FP}}, 
\label{Eq4}
\end{equation}
\begin{equation}
{\rm Recall = \frac{TP}{TP + FN}}, 
\label{Eq5}
\end{equation}
\begin{equation}
{\rm F1-score = 2 \times \frac{Precision \times Recall}{Precision + Recall}}, 
\label{Eq6}
\end{equation}
where TP (True Positives) is the number of positive samples correctly classified; TN (True Negatives) is the number of negative samples correctly classified; FP (False Positives) is the number of negative samples misclassified as positive; FN (False Negatives) is the number of positive samples misclassified as negative.

Accuracy, as shown in Equation~\ref{Eq3}, reflects the proportion of all correctly classified samples and provides a direct metric for evaluating overall model performance. However, this indicator may be ineffective in the case of imbalanced datasets. Consequently, it is primarily utilized to search for the optimal model on the validation dataset, rather than to evaluate the detailed performance of each class.

Precision, also known as Purity in astronomy, quantifies the fraction of true positives among the predicted positives. Recall represents the proportion of all positive samples that are successfully identified by the model, equivalent to Completeness in astronomy. It is important to note that ML practitioners often enhance classification confidence by setting a threshold for output probabilities, potentially excluding data below the threshold. When calculating Recall, some studies count these below-threshold samples as FN, while others completely ignore them \citep{Nakoneczny2021, Yang2023}. In scenarios where they are included, Recall still aligns with Completeness, otherwise, Recall needs to be adjusted by multiplying the fraction of samples above the threshold ($\rm Frac_{thre}$). Therefore, in our analysis of the testing dataset, we compute $\rm Frac_{thre}$ for different thresholds. 
The F1-score, the harmonic average of Precision and Recall, is particularly valuable as it provides a balanced measure of classification performance, reflecting the overall efficacy of the model across each category of sources. This metric was instrumental in evaluating the ability of our MNN to classify astronomical sources. Figure~\ref{fig3} displays the distribution of these evaluation metrics across different output probability thresholds.

Additionally, the confusion matrix is a powerful tool for evaluating classification problems, offering a visual representation of the relationship between model predictions for each class and the actual ground-truth categories. Figure~\ref{fig4} illustrates the results at the maximum output probability ($p_{\rm max}$), where each column represents the number of predictions for the corresponding class, and each row represents the ground-truth classes. To render the classification precision of each category more discernible, we have normalized the confusion matrix and displayed it in a percentage format.

\subsection{Implementation} \label{sec:3.3}
Our neural network model is implemented and trained using Keras\footnote{\url{https://keras.io}}, a high-level open-source API that provides a user-friendly interface built on top of the TensorFlow\footnote{\url{https://www.tensorflow.org}} framework. For the training and testing process, we utilized a GPU-accelerated computer equipped with NVIDIA GeForce RTX 3080 laptop GPU to enhance computational efficiency.

During the training phase, the hyperparameters such as batch size and regularization factor were determined through repeated experiments. We found that setting the batch size and L2 regularization factor to 128 and 0.001, respectively, was sufficient to achieve good performance on the testing dataset (containing 20,000 samples). To efficiently pinpoint the optimal model, we tested multiple learning rates ranging from 1e-3 to 1e-5.  The results revealed minimal performance differences among these rates, and we ultimately selected 1e-4 as the learning rate due to its stability and computational efficiency. Training was conducted for a total of 50 epochs, as the validation loss and accuracy were observed to plateau after approximately 20 epochs. The training and validation loss, along with the accuracy for each epoch, are presented in Figure \ref{fig5}. Each epoch took approximately 32 seconds, with a total training time of $\sim$27 minutes. We then selected the best model based on the minimum loss and maximum accuracy on the validation dataset. This model achieved nearly identical performance on the testing dataset, and we arbitrarily chose the model with the lowest loss to make predictions on the full dataset.

\begin{deluxetable*}{lcccccccccc}[!ht]
\tablewidth{\textwidth}
\tabletypesize{\scriptsize}
\tablecaption{Classification Performance for Different Methods.\label{tab1}}
\tablewidth{\textwidth}
\tablehead{
	\multicolumn{1}{c}{Method}&
    \multicolumn{3}{c}{Star}&
	\multicolumn{3}{c}{QSO}&
	\multicolumn{3}{c}{Galaxy}\\
    \colhead{}&
	\multicolumn{1}{c}{Precision}&
	\multicolumn{1}{c}{Recall}&
    \multicolumn{1}{c}{F1-score}&
    \multicolumn{1}{c}{Precision}&
	\multicolumn{1}{c}{Recall}&
    \multicolumn{1}{c}{F1-score}&
    \multicolumn{1}{c}{Precision}&
	\multicolumn{1}{c}{Recall}&
    \multicolumn{1}{c}{F1-score}
		}
\startdata   
MNN (Img1+Pho9) & 0.9826 & 0.9908 & 0.9867 & 0.9630 & 0.9325 & 0.9475 & 0.9896 & 0.9926 & 0.9911 \\ 
MNN (Img1+Pho4) & 0.9558 & 0.9309 & 0.9432 & 0.9003 & 0.8838 & 0.8920 & 0.9851 & 0.9944 & 0.9897 \\ 
MNN (Img4+Pho4) & 0.9543 & 0.9337 & 0.9439 & 0.9086 & 0.8810 & 0.8946 & 0.9854 & 0.9954 & 0.9903 \\ 
CNN (Img4) & 0.8982 & 0.8876 & 0.8929 & 0.8163 & 0.7859 & 0.8008 & 0.9816 & 0.9901 & 0.9858 \\ 
MPL (Pho9) & 0.9709 & 0.9730 & 0.9720 & 0.9344 & 0.8757 & 0.9041 & 0.9791 & 0.9889 & 0.9840 \\ 
XGBoost (Pho9) & 0.9854 & 0.9854 & 0.9854 & 0.9534 & 0.9152 & 0.9339 & 0.9865 & 0.9916 & 0.9891
\enddata
\tablecomments{Evaluation metrics (Precision, Recall, and F1-score) for stars, quasars, and galaxies using different classification methods. "Img" and "Pho" indicate the number of imaging bands and photometric bands, respectively, used during the training process.}
\end{deluxetable*}

\section{Results} \label{sec:4}
We assessed the overall classification capability and applicability of the proposed MNN by examining its performance on an independent internal testing dataset. Furthermore, two external datasets were utilized to validate the robustness and generalization ability of the model beyond the training sample space. Finally, we applied the trained network to the full KiDS DR5 dataset and generated a new Star-QSO-Galaxy catalog for objects brighter than 23 mag in $r$-band.

\subsection{Performance on Testing Datasets} \label{sec:4.1}
\subsubsection{Internal Testing Dataset}\label{sec:4.1.1}
Although the validation dataset was not directly involved in model training, its role in model selection may inadvertently lead to overfitting, thereby increasing the risk of information leakage. To address this concern and objectively assess the performance of our model, we randomly selected 20,000 spectroscopically confirmed sources from the SDSS-KiDS cross-matched sample to compose an independent testing dataset. The sample was exclusively for final testing, without participating in the training or validation stages. This approach provides a more objective assessment by mitigating biases introduced during the training process and model selection.

On this isolated testing dataset, our MNN achieved an overall accuracy of 98.76\%, closely aligning with the 98.69\% accuracy observed in the validation dataset. Such consistency indicates that our model generalizes well from the validation phase to unseen new data, which is crucial for practical applications. We then conducted more detailed performance assessments on each class of sources, including Precision, Recall, and F1-score. The results are presented in Figure~\ref{fig3} and Table~\ref{tab1}.

The classification metrics revealed that galaxies exhibited the best performance with all three metrics exceeding 99\%. This success can be attributed to the large training sample size and the distinct morphological features of galaxies. This was followed by the stars, with all metrics also exceeding 98.6\%, reflecting robust identification of these two types of celestial objects. QSOs achieved slightly lower Precision and Recall at 96.71\% and 94.53\% respectively, likely due to the relatively smaller training sample size, as well as the potential for misclassification of some low-redshift active galactic nuclei (AGNs) as galaxies due to contamination from their host galaxies. This latter point is evidenced by the confusion matrix depicted in Figure~\ref{fig4}, where the majority of misclassified QSOs were grouped with galaxies. Despite this, a robust F1-score of 95.6\% demonstrates that the model is still reliable in terms of QSO classification.

In ML paradigms, our model assigns a probability vector to each target for classification, with each value representing the likelihood of a predefined class (such as Stars, QSOs, or Galaxies in our definition). The sum of these probabilities is 1, and typically, a higher probability (p) for a target being classified into a certain class indicates greater confidence, which implies higher Purity (Precision). Therefore, we can set a threshold for $p$, retaining only those samples that exceed this threshold as the final result, thereby enhancing the sample Purity. It is important to note that this strategy, while beneficial for sample Purity, may decrease the Completeness of the sample. For example, a higher threshold may exclude objects with insufficient classification confidence, potentially leading to the absence of some actual category members in the final catalog.

To balance Purity and Completeness in various astronomical applications, we tested the performance of our MNN at different thresholds. Figure~\ref{fig3} shows the variations in evaluation metrics as a function of the threshold. As expected, an increase in the threshold can improve Precision, Recall, and the F1-score, whereas $\rm Frac_{thre}$ and the derived Completeness ($\rm Frac_{thre} \times$ Recall) exhibit a declining trend. This trend suggests that raising the threshold for $p$ indeed amplifies the Purity of the sample but at the cost of Completeness.

An analysis of Figures~\ref{fig3} reveal that the choice of probability threshold significantly impacts the classification of QSOs, in contrast to a minor effect on the classification of Galaxies. This is primarily because the model already achieves excellent performance in predicting Galaxies, leaving limited scope for further refinement. Intriguingly, we found that when the threshold is fixed at 0.9, the Completeness for all categories can be maintained above 90\%, and the Purity exceeds 98\%. As the threshold is further increased to 0.99, Purity can reach above 99\%, while Completeness still maintains a level of 82\%. Beyond this point, Completeness quickly declines despite a marginal gain in Purity. Hence, setting the probability threshold between 0.9 to 0.99 appears to offer an effective balance between Purity and Completeness.

\subsubsection{External Testing Datasets} \label{sec:4.1.2}
Considering the issue of pre-selected sources in the SDSS dataset, there is potential for our neural network model to overfit to this data. To comprehensively evaluate the generalization capability of the model in practical applications, it is necessary to validate it on external datasets. Currently, most publicly released spectroscopic survey data have limited overlap with KiDS, which constrains the availability of extensive testing samples. We primarily utilize Gaia and GAMA as external data sources to verify the generalization performance of our model on stellar and galaxy classification, respectively. For QSOs, while GAMA provides partial coverage of high-redshift samples, a uniformly selected and sufficiently large pure QSO dataset is currently lacking. Therefore, instead of relying on a comprehensive spectroscopic validation dataset, we assess the effectiveness of the model in QSO identification by comparing the number of QSO candidates it recovers across the survey footprint.

Most targets in the Gaia database do not have corresponding spectroscopic observations, but its high-precision astrometric measurements enable us to filter out a pure stellar sample. This sample is considered unbiased in terms of stellar radiation and can effectively avoid potential selection effects. Through cross-matching with KiDS, we obtained over 3.4 million independent testing samples (see Section~\ref{sec:2.3}). None of these data were involved in model tuning or evaluation. The test results showed that as high as 99.74\% of the targets were correctly predicted as stars, even exceeding the performance on the internal testing dataset. This discrepancy may be attributed to the contamination present in SDSS spectroscopic classifications (see Section \ref{sec:5.2}) and the fact that GAIA dataset includes brighter sources, which could hint at the superior performance of our algorithm in star classification under these conditions.

\begin{figure}[!ht]
\centering
\includegraphics[scale=0.55]{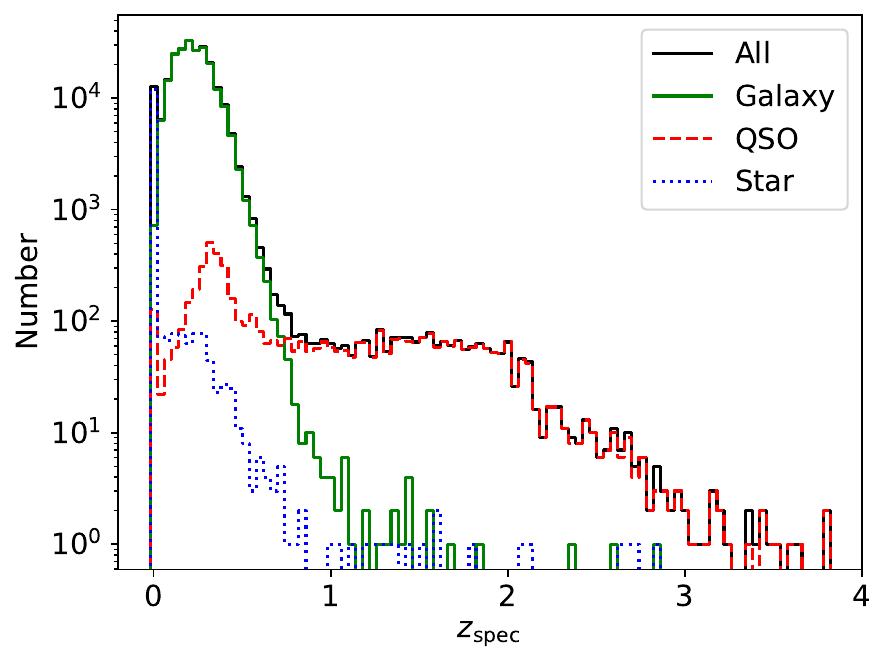}
\caption{Redshift distribution for the GAMA testing dataset. The black solid line shows the distribution for all targets, while the blue dotted line, red dashed line, and green solid line represent the distributions for targets predicted as stars, quasars, and galaxies by the MNN, respectively.} 
\label{fig6}
\end{figure}

For galaxies, we adopted the abundant spectroscopic redshift galaxy sample from GAMA for testing. Over 230k of these targets that overlap with KiDS but not with SDSS. The prediction results revealed that only 97.70\% of the targets were categorized as galaxies, with the remaining 1.21\% classified as stars and 1.11\% as quasars, which did not appear to meet expectations. However, upon further examination of the redshift distribution of these targets (see Figure~\ref{fig6}), we found that the majority of those predicted as stars had redshifts around 0, confirming that they indeed reside within the Milky Way. After removing these low-redshift (arbitrarily chosen $z <$ 0.04) targets, the fraction of categorized as galaxies increased to 98.72\%, while stars and quasars accounted for 0.26\% and 1.02\%, respectively. Figure~\ref{fig6} also shows that targets with redshifts greater than 0.8 were almost all classified as QSOs. Given the depth of the GAMA survey, it is unlikely to observe such high-redshift galaxies, so these classification results should be reasonable. If QSO classification is also regarded as correct, we can conclude that the classification Recall for extragalactic objects reached 99.74\%.

To estimate the number of QSOs expected in the KiDS footprint under a given apparent magnitude limit, we adopt the method described in Section \ref{sec:2.2} to select QSOs in the Stripe 82 region of the SDSS\footnote{We use the subregion defined by $32^{\circ} < {\rm R.A.} < 45^{\circ}$ and $-1\fdg25 < {\rm Decl.} < 1\fdg25$, where the QSO surface density is highest. We select all spectroscopically confirmed QSOs in this region (excluding duplicate observations), and compute the QSO surface density as a function of apparent magnitude. To obtain a robust estimate, we randomly sample 100 independent 4 deg$^{2}$ subregions and adopt the median surface density as the final value.}, where spectroscopic completeness is relatively high \citep{Yang2023}. At $r < 21$, the QSO surface density derived from SDSS is $\sim$83 deg$^{-2}$, corresponding to $\sim$112.5k QSOs over the 1350 deg$^{2}$ KiDS footprint. Our MNN model predicts a comparable number of $\sim$119.5k QSO candidates, indicating good agreement. At $r < 22$, the SDSS data yield an estimated $\sim$209.5k QSOs, while our model identifies a significantly higher number of $\sim$394.5k—nearly a factor of two larger. This discrepancy likely reflects the rapid decline in SDSS spectroscopic completeness at the faint end.

We also adopt quasar luminosity functions (QLFs) to estimate the number of QSOs with $r < 22$\footnote{We assume a power-law continuum for QSOs, $f_{\nu} \propto \nu^{\alpha_\nu}$, with a spectral index of $\alpha_\nu = 0.5$, and convert $r$-band apparent magnitudes to absolute magnitudes at 1450 \AA. The QLF is integrated down to $M_{1450} = -19$ to include faint AGNs.}. Using the pure luminosity-evolution and luminosity and density evolution QLFs from \citet{PalanqueDelabrouille2016}, we estimate a total of $\sim$226.8k QSOs over the KiDS area, which is consistent with the estimate from Stripe 82. In comparison, the QLF from \citet{Liu2022} yields a higher estimate of 584.2k, slightly exceeding our prediction. This suggests that the number of QSOs predicted by our MNN model remains within the plausible range expected from QLFs, further supporting the reliability of our selection method. For sources fainter than 22.3 mag, however, the MNN model predicts significantly more QSOs than the QLFs (e.g., only 788.7k QSOs are predicted by the QLF for $r < 23$). This discrepancy may reflect increasing uncertainties in the QLFs at the faint end, or limitations of the DNN model when extrapolating beyond the magnitude range of the training sample.

In the above test, we did not impose any restrictions on the threshold of $p$, but the model still demonstrated excellent performance for both Galactic and extragalactic targets. This result supports the stability and reliability of our algorithm, at least for bright-end targets. \citet{Nakoneczny2021} employed the XGBoost method to construct a reliable QSO sample from KiDS DR4, limited to sources with $r < 22$. We cross-matched their catalog with our predictions and found that 79\% of their QSOs were also classified as QSOs by our model. Most of the mismatched objects have $p < 0.9$, and after removing them, the disagreement rate drops to $\sim$6\%. These findings establish a solid foundation for future applications to more extensive samples and other survey data.

\begin{deluxetable*}{cccccccccccc}[!ht]
\tablewidth{\textwidth}
\tabletypesize{\scriptsize}
\tablecaption{Classification Performance at Different Magnitude Ranges.\label{tab2}}
\tablewidth{\textwidth}
\tablehead{
	\multicolumn{1}{c}{Train Range}&
    \multicolumn{1}{c}{Test Range}&
    \multicolumn{3}{c}{Star}&
	\multicolumn{3}{c}{QSO}&
	\multicolumn{3}{c}{Galaxy}\\
    \colhead{}&
    \colhead{}&
	\multicolumn{1}{c}{Precision}&
	\multicolumn{1}{c}{Recall}&
    \multicolumn{1}{c}{F1-score}&
    \multicolumn{1}{c}{Precision}&
	\multicolumn{1}{c}{Recall}&
    \multicolumn{1}{c}{F1-score}&
    \multicolumn{1}{c}{Precision}&
	\multicolumn{1}{c}{Recall}&
    \multicolumn{1}{c}{F1-score}
		}
\startdata   
$r < 21$ & $21 < r \leqslant 22$ & 0.9237 & 0.9685 & 0.9456 & 0.9552 & 0.8723 & 0.9118 & 0.9806 & 0.9910 & 0.9858 \\ 
$r < 20$ & $20 < r \leqslant 21$ & 0.9647 & 0.9799 & 0.9722 & 0.9064 & 0.9177 & 0.9120 & 0.9851 & 0.9792 & 0.9822 \\ 
$r < 19$ & $19 < r \leqslant 20$ & 0.9634 & 0.9820 & 0.9726 & 0.9685 & 0.9036 & 0.9349 & 0.9684 & 0.9869 & 0.9776 \\ 
$r < 18$ & $18 < r \leqslant 19$ & 0.9838 & 0.9861 & 0.9849 & 0.9711 & 0.8837 & 0.9253 & 0.9835 & 0.9949 & 0.9892 \\ 
$19 < r \leqslant 21$ & $21 < r \leqslant 22$ & 0.9231 & 0.9638 & 0.9430 & 0.9444 & 0.8797 & 0.9109 & 0.9822 & 0.9895 & 0.9858 \\ 
$18 < r \leqslant 20$ & $20 < r \leqslant 21$ & 0.9659 & 0.9609 & 0.9634 & 0.8814 & 0.9207 & 0.9006 & 0.9848 & 0.9767 & 0.9807 \\ 
$17.5 < r \leqslant 19.5$ & $20.5 < r \leqslant 20.5$ & 0.9638 & 0.9820 & 0.9728 & 0.9472 & 0.9305 & 0.9387 & 0.9803 & 0.9807 & 0.9805 \\ 
$17 < r \leqslant 19$ & $19 < r \leqslant 20$ & 0.9231 & 0.9638 & 0.9430 & 0.9444 & 0.8797 & 0.9109 & 0.9822 & 0.9895 & 0.9858 
\enddata
\tablecomments{Evaluation metrics (Precision, Recall, and F1-score) for stars, quasars, and galaxies trained on datasets with different $r$-band magnitude ranges.}
\end{deluxetable*}

\subsection{Testing the Extrapolation Performance of the Model} \label{sec:4.2}
Due to the inherent extrapolation uncertainty of ML models, their applicability typically depends on the coverage of the feature space by the known data. In our study, the SDSS spectroscopic depth in the $r$-band is approximately 22 mag, which is $\sim$2 mag brighter than KiDS, and there are also differences in the color distribution between them (see Figure~\ref{fig1}). Therefore, it is necessary to filter the prediction data to ensure the reliability of the final results. This requires understanding the specific feature space boundaries of the model, which is a challenge for neural network models. We primarily compared the brightness and color distributions of the two databases—historically the most commonly used features—and found that the most significant difference was in brightness. By limiting our focus to targets with $r$-band magnitudes brighter than 23, we found that the color distribution of most KiDS data aligned with SDSS (e.g., 99.96\% of $u - g$ colors were within the range of -1 to 4). This phenomenon may be related to the requirements in the data preprocessing stage, where all bands must have reliable photometric detection values, thus excluding some extremely red and blue targets. The color consistency increases our confidence in the prediction results for brighter sources, but also introduces limitations for the model in identifying targets with extreme color properties.

To evaluate the applicable brightness range of the model, we divided the cross-matched SDSS data into multiple subsets based on their $r$-band magnitudes and trained the model on each subset to assess its extrapolation performance. The training (and validation) datasets were constructed using two different approaches:
\begin{enumerate}
\item Threshold-based division: Four brightness thresholds were defined, and for each threshold, all sources brighter than the given value were grouped into a single training dataset.

\item Bin-based division: The data were divided into four bins with equal magnitude intervals, and all sources within each bin were used as an independent training dataset.
\end{enumerate}
For each trained model, we evaluated its performance using datasets fainter than the corresponding training dataset by a series of thresholds ranging from 0.5 to 1.5 mag. The results, summarized in Table \ref{tab2}, indicate that the model maintains acceptable performance when extrapolating up to 1.0 mag beyond the training range. For example, the F1-scores for stars, QSOs, and galaxies consistently exceed 94\%, 90\%, and 97\%, respectively, across all tests. This suggests that our MNN has the potential to handle targets with magnitudes up to $\sim$23 mag. However, additional testing is required to fully characterize its extrapolation limitations. Therefore, sources with $r$-band magnitudes brighter than 22 mag are regarded as reliable targets, while those with magnitudes between 22 and 23 mag are treated as acceptable extrapolated targets.

\begin{figure*}[!ht]
\centering
\includegraphics[scale=0.81]{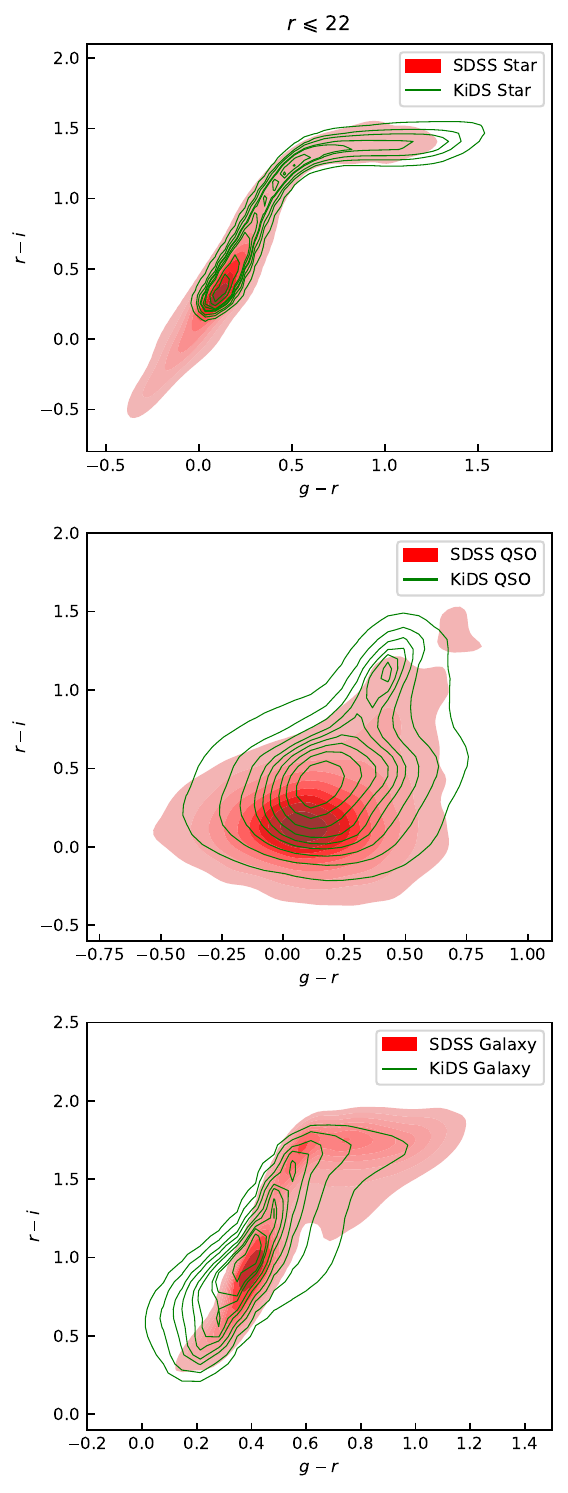}
\includegraphics[scale=0.81]{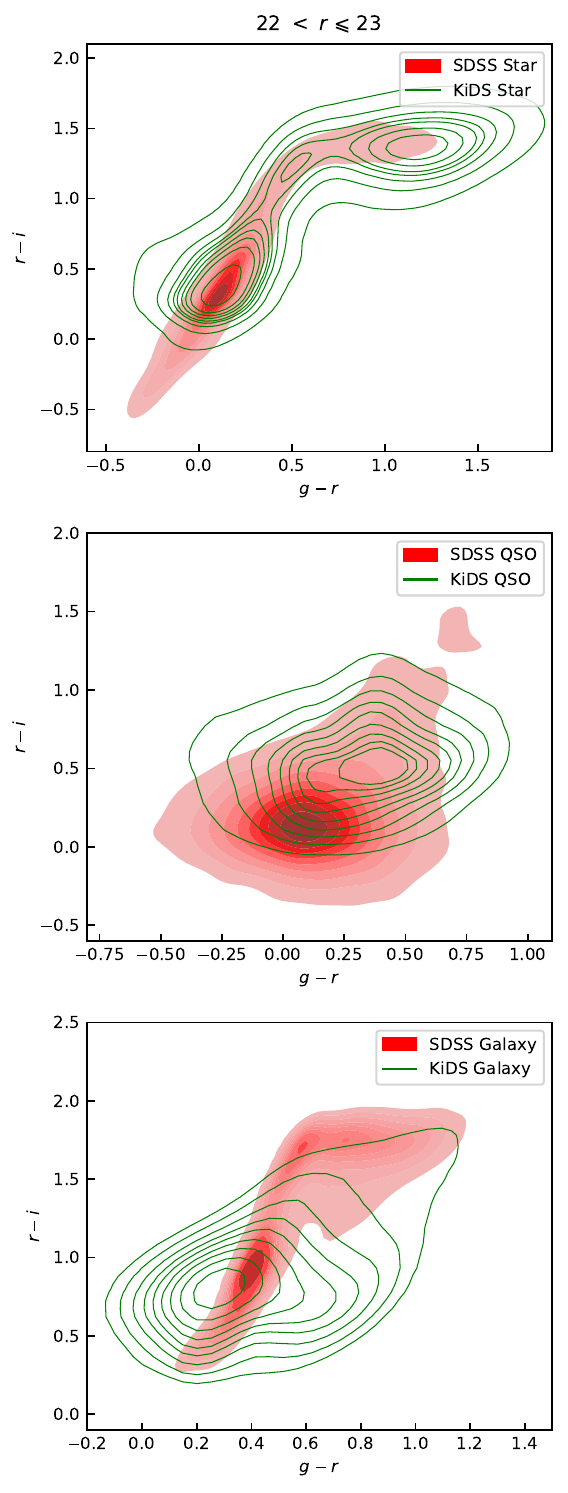}
\caption{Color-color diagrams ($g - r$ vs. $r - i$) for KiDS and SDSS datasets. From top to bottom, the panels represent stars, QSOs, and galaxies. The left panels show objects with $r \leqslant 22$, while the right panels correspond to objects with $22 < r \leqslant 23$. The red shaded regions indicate the cross-matched data between SDSS and KiDS, and the green contours represent all KiDS data within the corresponding magnitude range.} 
\label{fig7}
\end{figure*}

\begin{figure}[!ht]
\centering
\includegraphics[scale=0.38]{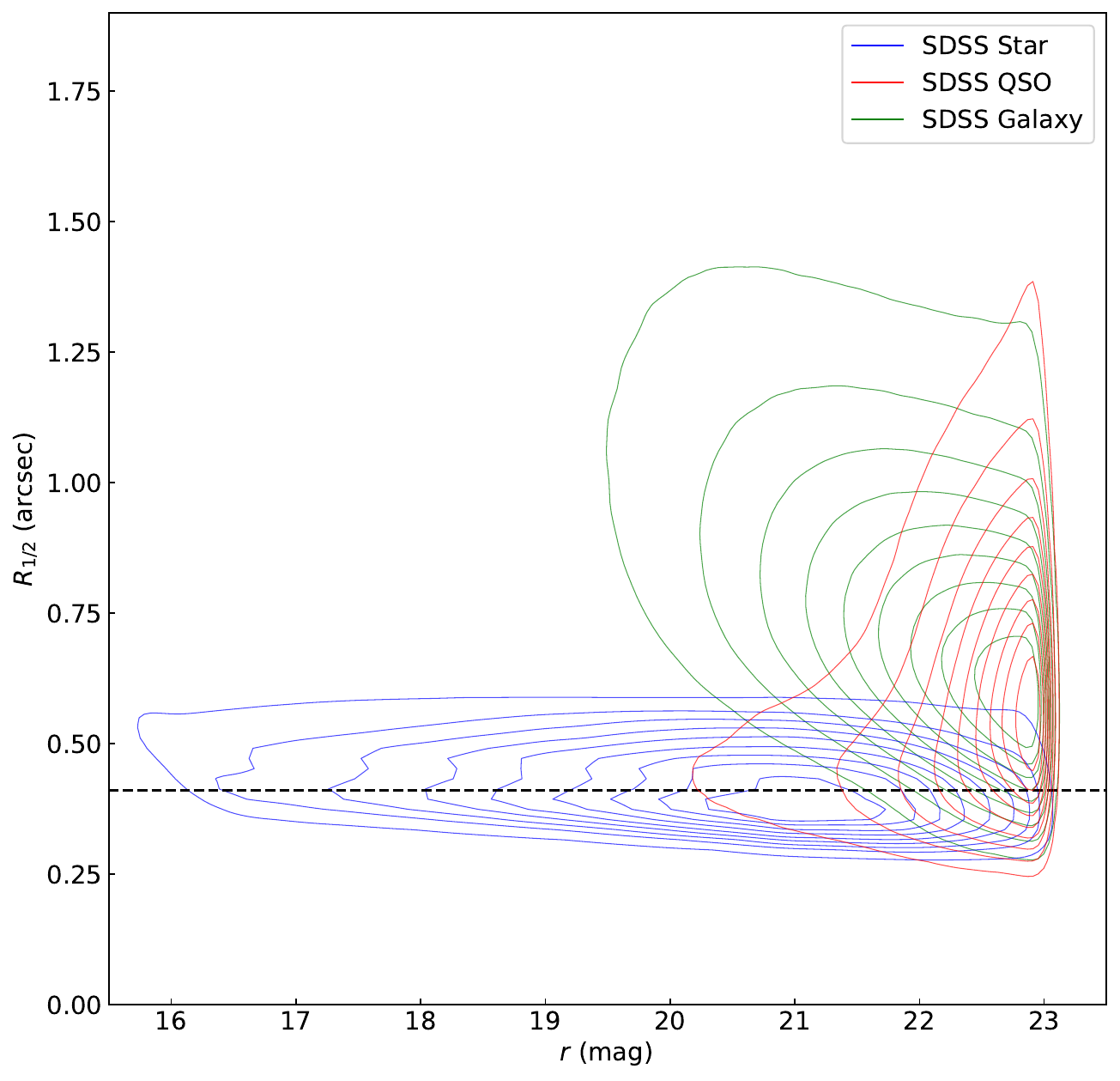}
\caption{The $r$ vs. $R_{1/2}$ for objects in our catalog. The contours represent the density distribution of stars (blue), QSOs (red), and galaxies (green). The dashed line (black) at $0.41^{\prime\prime}$ indicates the $R_{1/2}$ corresponding to the average KiDS $r$-band seeing of $0.7^{\prime\prime}$, assuming a Gaussian PSF where $R_{1/2} \approx 0.5887 \times \mathrm{FWHM}$.} 
\label{fig8}
\end{figure}

\begin{figure}[!ht]
\centering
\includegraphics[scale=0.7]{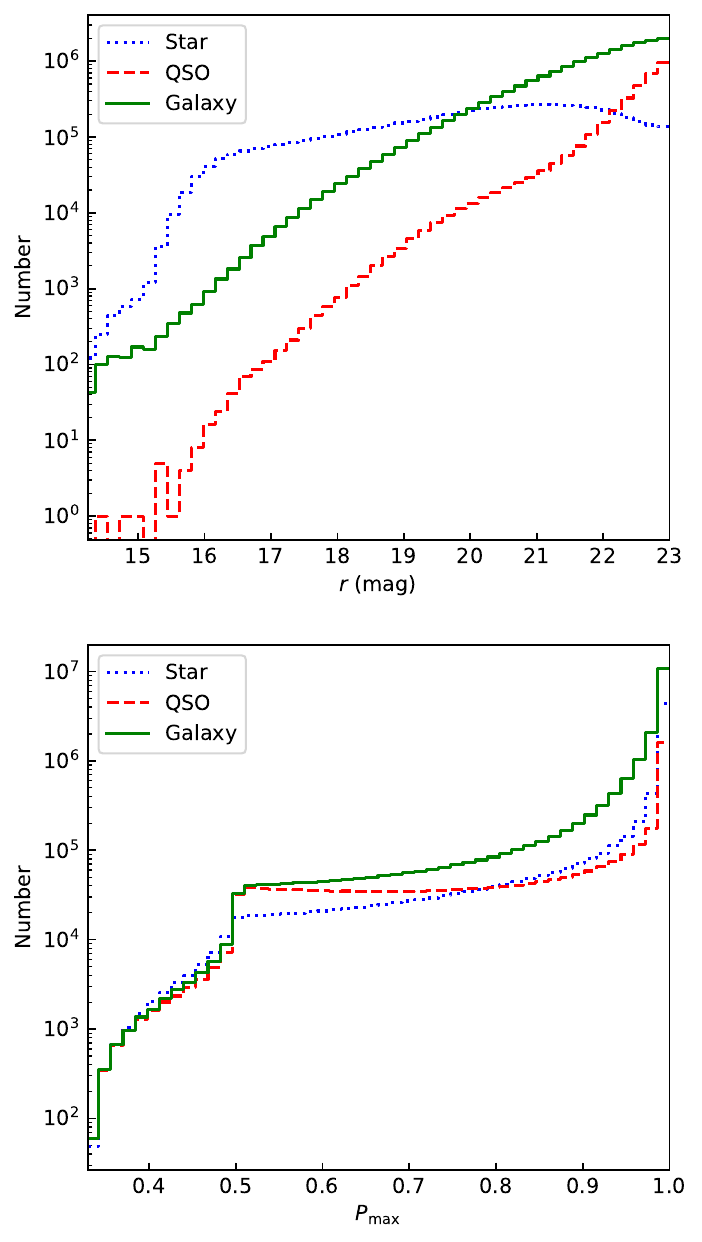}
\caption{Distribution of $r$-band magnitude (top panel) and maximum output probability (bottom panel) for each type of object in the value-added catalog. Each panel plots the distribution for targets predicted as stars (blue dotted line), quasars (red dashed line), and galaxies (green solid line).} 
\label{fig9}
\end{figure}

\begin{figure*}[!ht]
\centering
\includegraphics[scale=0.5]{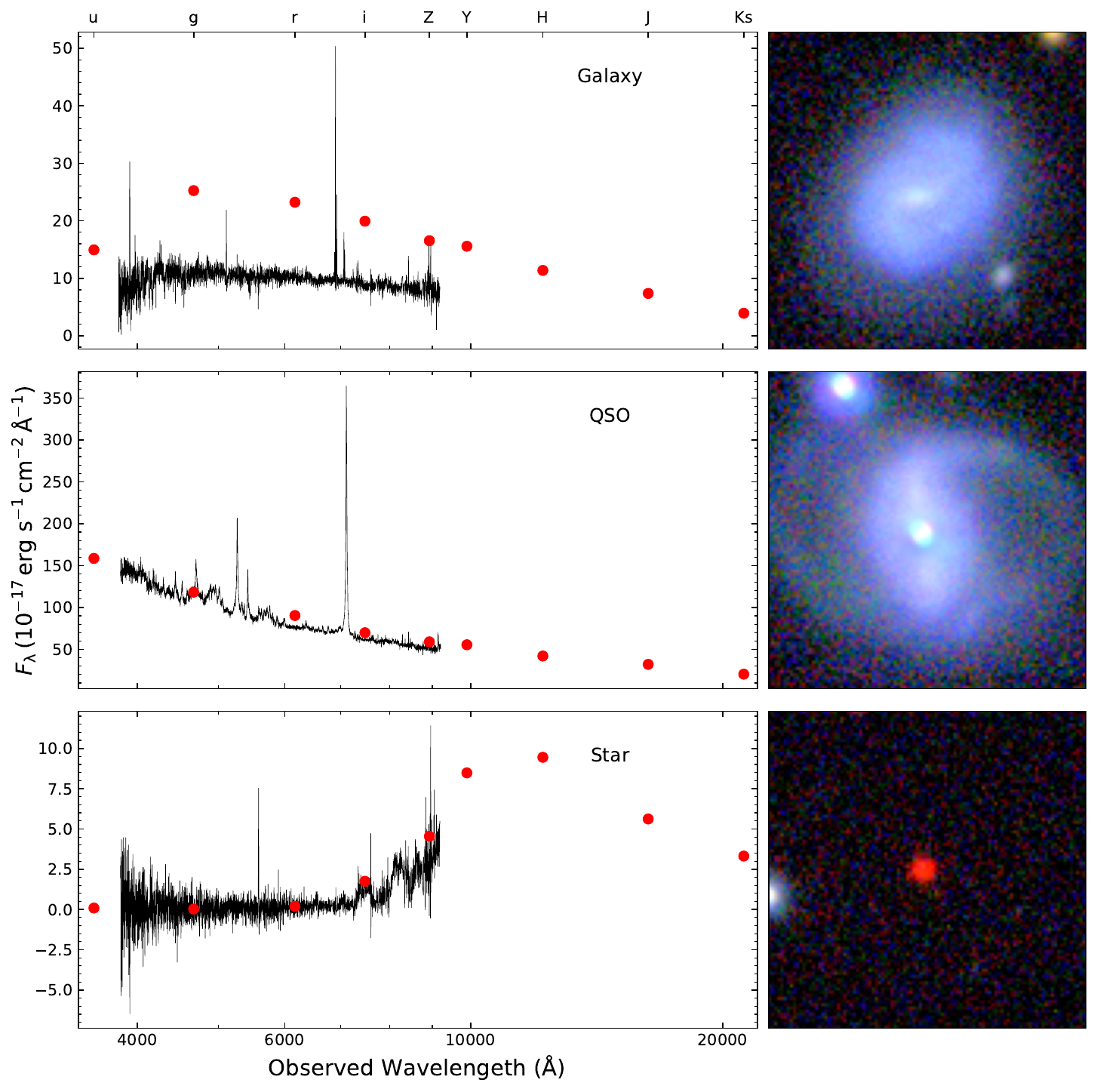}
\caption{Examples of spectra, SEDs, and images for different categories. The left panels show the SDSS spectra (black solid line) and the SEDs from KiDS nine bands (red points). The SEDs agree well with the spectra for star and QSO, while galaxy photometric points tend to be higher due to the larger KiDS aperture compared to the SDSS fiber. The right panels show color-composited images from KiDS $gri$-bands.} 
\label{fig10}
\end{figure*}

\begin{deluxetable*}{ccccccccc}[!ht]
\tablecaption{Column Descriptions of the KiDS DR5 Galaxy-QSO-Star Catalog\label{tab3}}
\tablewidth{0pt}
\tablehead{
    \colhead{No.}& 
    \colhead{Column Name}& 
    \colhead{Description}& 
    \colhead{Units}& 
    \colhead{Datatype}
        }
\startdata
1 & Name & KiDS unique source identifier from ASTRO$_{\rm band}$ table & - & string \\
2 & RA & J2000 Right Ascension & degree & double \\
3 & DEC & J2000 Declination & degree & double \\
4 & $r$ & KiDS $r$-band GAaP magnitude & mag & double \\
5 & class & classification: STAR/QSO/GALAXY & - & string \\
6 & $p_{\rm max}$ & Maximum output probability & - & double
\enddata
\tablecomments{Description of columns in the KiDS DR5 Galaxy-QSO-Star Catalog. The table lists the properties stored for each object, including identifiers, coordinates, magnitudes, classification results, and probabilities. (The full catalog can be accessed at \url{https://cosviewer.com/datasets/kids-dr5-target-classify}. The $r$-band magnitudes will be updated after the photometry data from KiDS DR5 is fully released.)}
\end{deluxetable*}

\subsection{Application to KiDS DR5} \label{sec:4.3}
We apply the trained MNN model to the entire KiDS DR5 dataset to predict the categories of target sources. Our goal is to compile a new Star-QSO-Galaxy catalog covering an area of approximately 1350 deg$^{2}$. According to our prior testing outcomes, we only classify targets with an $r$-band magnitude less than 23. To facilitate diverse research interests, we provide the $r$-band magnitude and the associated $p_{\rm max}$ for each target, with detailed information listed in Table~\ref{tab3}.

This value-added catalog contains a total of 27,335,836 targets, of which 6,549,300 are identified as stars, 2,804,654 as quasars, and 17,981,882 as galaxies. Notably, 13,068,547 targets fall within the $22 < r \leqslant 23$ mag range in the $r$-band, which are extrapolated results, accounting for nearly half ($\sim$47.81\%) of the total number of targets. This portion of the data exceeds the brightness range directly trained by our model, yet the model has demonstrated acceptable accuracy and stability within this magnitude range in previous tests. Therefore, we decide to include these extrapolated results in the final catalog.

To validate the effectiveness of our final classification results, we compared the $g - r$ vs. $r - i$ color-color distributions of stars, QSOs, and galaxies in our KiDS catalog with those in the SDSS dataset. As shown in Figure~\ref{fig7}, the reliable targets exhibit good agreement with the SDSS dataset, whereas the extrapolated targets display some deviations but remain largely consistent. This demonstrates that our model effectively captures the overall color properties of different sources. Furthermore, we examined the size-magnitude diagram for each class of sources for each class of sources, which is a commonly used method to differentiate between stars and galaxies \citep[e.g.,][]{Zuntz2018}. Here, we adopt the half-light radius ($R_{1/2}$) to represent size, and the magnitude is in the $r$-band. The results are shown in Figure~\ref{fig8}. It is evident that stars are primarily located at the bright end, with sizes typically consistent with the PSF, reflecting their point-source nature. In contrast, galaxies are predominantly found at the fainter end and exhibit significantly larger sizes compared to stars, indicative of their more extended structures. Quasars are primarily distributed between the star and galaxy populations in the size-magnitude space. This distribution aligns well with known observational characteristics and supports our classification results.

Figure~\ref{fig9} presents the $r$-band magnitude distribution for each category, revealing a dominance of stars at the brighter end ($r < 20$ mag), while the number of quasars and galaxies increases rapidly towards the faint end, which aligns with empirical expectations. Moreover, the distribution of $p_{\rm max}$ (Figure~\ref{fig9}) indicates exceedingly high classification confidence for the majority of targets. Over 85\% of the catalog entries have a $p_{\rm max} \geqslant 0.9$, and remarkably, for targets with $r \leqslant 22$ mag, this proportion exceeds 91\%. This means that even under relatively stringent selection criteria, we can still obtain about 13 million high-confidence classification results.

\section{Discussion} \label{sec:5}
\subsection{Model Interpretation} \label{sec:5.1}
Neural network models are often regarded as ``black boxes'' with internal decision-making processes that are opaque to humans. This opacity makes it difficult to directly analyze the intrinsic connections between celestial object classification and their physical properties. Here, we attempt to use empirical knowledge to explain the features that the model may have learned during its training, aiming to better understand the physical mechanisms behind its excellent performance.

The proposed MNN extracts information from two input branches: SED and image. As shown in Figure \ref{fig10}, different types of celestial objects inherently exhibit distinct SED and morphological characteristics. The SED primarily reflects the differences in internal radiation mechanisms and the distances (redshifts) of the objects. Stars, for instance, typically exhibit SEDs that can be approximated by blackbody radiation, with peak positions mainly determined by their effective temperatures. On the other hand, the spectrum of quasars in the ultraviolet to optical bands is dominated by the ``big blue bump'', a feature produced by high-temperature radiation from the accretion disk, while the near-infrared is predominantly emission from hot dust \citep{Sanders1989, Elvis1994}. The multi-wavelength luminosity of galaxies is more complex, as it involves the combined effects of stellar populations and dust \citep{Conroy2013}. The nine-band photometric data provided by KiDS spans a broader wavelength coverage than SDSS spectra, potentially yielding additional physical information. For example, the SED peak of M-type stars is located in the near-infrared band (see Figure \ref{fig10}), and as the redshift increases, the contribution of the ``big blue bump'' in quasars becomes more pronounced in the optical-near-infrared bands. This prior knowledge suggests that the model may achieve classification by learning the relative brightness across multiple bands. 

For the imaging side, it can provide rich details of the morphology of celestial objects. These morphological features not only contain information about the physical nature of objects but might also refine flux measurements, as the shapes of different objects can affect the results of GAaP measurements. To test the impact of images on classification, we separately used MLP to classify the SED (see Appendix \ref{sec:appendixA}), and found that the performance was significantly inferior to the MNN. This demonstrates that incorporating morphological information can indeed significantly improve classification efficacy. To further confirm which information directly aids classification, we normalized all images and found that there was a negligible effect on model performance. This implies that during the classification process, the CNN branch primarily relies on the physical morphological features rather than simply re-measuring the flux from the images.

Moreover, if images provide more accurate flux measurements, directly utilizing images from all bands should yield more precise SED results. To investigate this, we conducted the following tests using all $ugri$ images released by KiDS (for a fair comparison, we also utilized the corresponding 4-band photometric data):
\begin{enumerate}
\item We retrained the MNN shown in Figure \ref{fig2}, but the SED branch only used photometric data from the $ugri$ bands (along with colors from 3 adjacent bands).

\item We removed the SED branch from Figure \ref{fig2} and added 3 additional CNN branches (resulting in 4 CNN branches in total), with each branch taking image data from a single band as input. The model was then trained directly using images from all $ugri$ bands.

\item Based on the model with 4-branch CNN neural network, we added the SED branch for joint training.
\end{enumerate}
The test results are listed in Table \ref{tab1}. The model trained using images from all four bands performed worse than the MNN, indicating that images may not provide more accurate flux measurements than the photometric data. When considering both SED and images, increasing the number of input images did not lead to significant performance improvements. This suggests that images do not play a crucial role in correcting photometric accuracy, as using the image from the same band should provide more accurate corrections for a specific band. These findings also demonstrates the efficiency and practicality of the multimodal approach, which achieves comparable performance using only a single-band image combined with a few photometric data points. This significantly reduces the computational and storage resources required, compared to models that rely on processing multiple image bands.

This is consistent with the human inferential strategies, where we prefer to use the physical properties of morphology itself rather than mere flux correction when classifying. For example, extended galaxies are easily distinguishable from stellar point sources, and the size can be used to estimate the distance of the galaxy (see also Figure~\ref{fig8}). For high-redshift quasars and compact galaxies, they may appear more like stars in images, but we can at least rule out the possibility of them being low-redshift extragalactic objects.

In summary, the MNN can extract luminosity information from photometric data and morphological features from image data, and mapping both to a high-dimensional feature space via a nonlinear function. Within this space, various categories are effectively separated, thus achieving the purpose of classification. The accuracy and efficiency of this method hold promise for widespread application in next-generation multi-color survey projects such as LSST, WFST \citep{Wang2023}, and Mephisto \citep{Han2022}. Future space survey projects (e.g., CSST and Euclid) will provide unprecedented high-resolution images and slitless spectral data, which will further refine our model and enhance the precision of astronomical classification. Moreover, the anticipated wealth of spectral data from future spectroscopic surveys will enable our model to extend its performance to fainter magnitudes.

\begin{figure*}[!ht]
\centering
\includegraphics[scale=0.48]{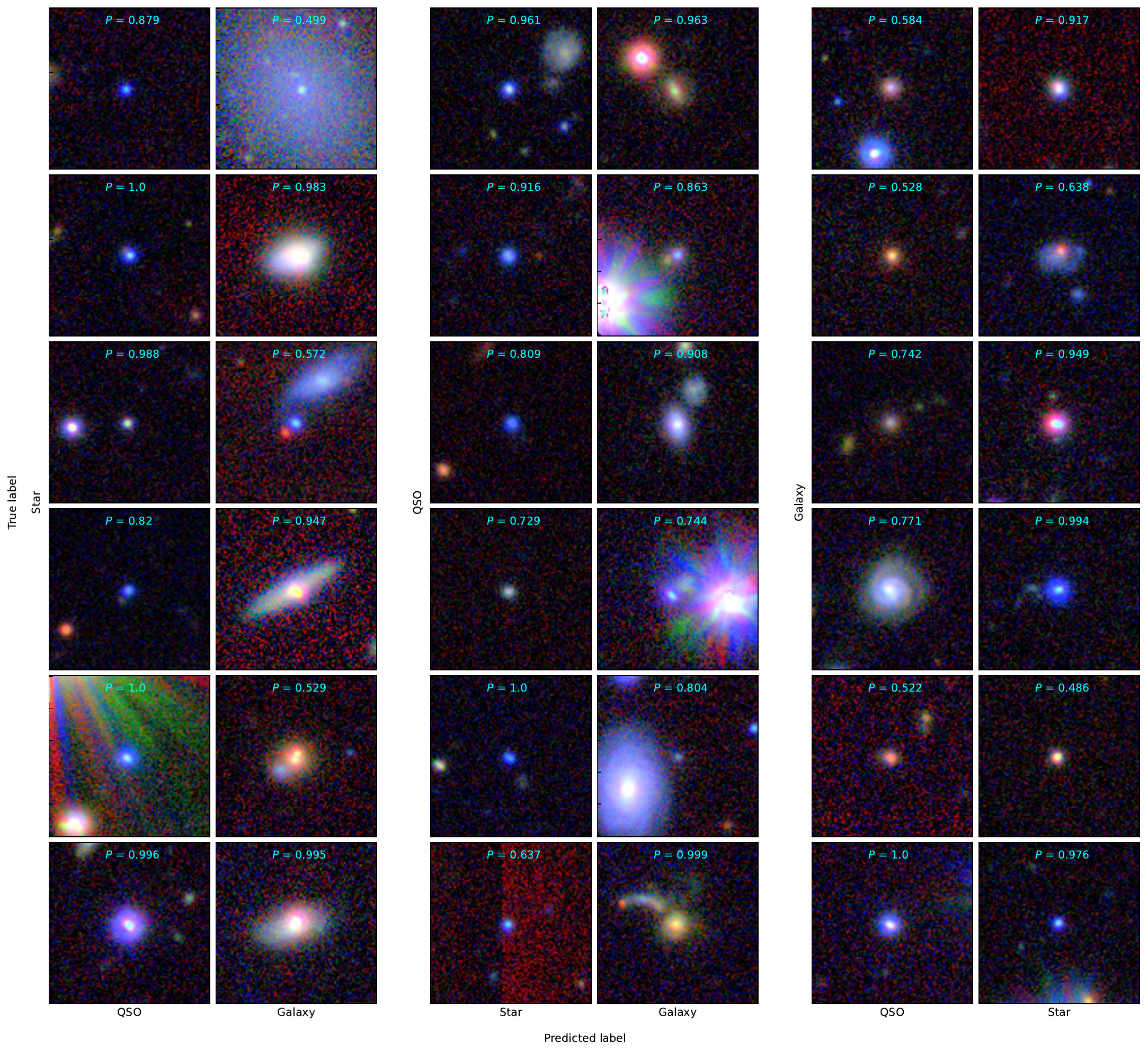}
\caption{$gri$ color-composited images of some misclassified targets. The vertical axis represents the spectral classification labels from SDSS, and the horizontal axis represents the predictions made by the MNN.} 
\label{fig11}
\end{figure*}

\subsection{Limitations and Challenges} \label{sec:5.2}
Several studies have shown that tree-based ML methods, such as XGBoost \citep{Chen2016} or Random Forest \citep{Breiman2001}, often outperform other algorithms when performing classification tasks on structured data like photometry and colors \citep[e.g.,][]{ Zeraatgari2024, Nakoneczny2019}. To investigate this, we trained the widely used XGBoost algorithm on the same training data as our MNN. The results indicate that XGBoost indeed performs better than MLP, but falls short of our MNN (Table \ref{tab1}). This can be primarily attributed to the absence of morphological information in the training process. We note that \citet{Nakoneczny2021} achieved comparable performance to our method by incorporating structured morphological information \citep[stellarity index,][]{Bertin1996} into the XGBoost model. However, one of our goals is to develop a flexible and versatile approach that can directly extract morphological information from images. This approach not only enables real-time classification of image data but also captures the high-dimensional, non-linear physical properties underlying the data \citep{Lecun2015}. DNNs are particularly well-suited for this task due to advantages such as their ability to automatically extract complex and hierarchical features directly from raw data, as well as their effectiveness in modeling high-dimensional and nonlinear relationships.

In addition to the choice of ML algorithms, data quality fundamentally determines the upper limit of model performance. Clean data allows the model to focus on learning the underlying patterns that are consistent with the real-world problems during the training process, thereby enhancing its stability and interpretability. Conversely, dirty data may mislead the learning process of model parameters and evaluation results, increasing the risk of overfitting. Indeed, there are some misclassified targets in the SDSS spectral data \citep{Bolton2012, Lyke2020}. However, since the proportion of these misclassifications is relatively low (about only a few percent), the model still performs impressively well on the testing dataset. This can also be attributed to the hierarchical feature extraction of DNNs, which allows them to reduce the impact of random noise during the training process, as well as the use of techniques such as regularization and batch training \citep{Zhang2016, Goodfellow2016}. Through random inspection of misclassified samples, we found that some SDSS labels themselves were erroneous, and our MNN model was even able to effectively correct these errors, especially for galaxies misclassified as stars by SDSS (as shown in Figure~\ref{fig11}).

The performance of the model is also influenced by the data volume. A larger dataset can provide richer and more detailed feature representations, which helps the model learn complex patterns and effectively mitigate noise and outlier. This, in turn, enhances its accuracy and generalization ability. In our training, the number of quasar samples was significantly less than that of stars and galaxies, which somewhat limited the recognition capabilities of the model for these targets. To better understand the impact of training data volume on classification performance, we conducted a test in Appendix \ref{sec:appendixB} by artificially selecting subsets of the training data. The results show that the performance of all three source types improves as the training data volume increases, with quasars showing the most significant improvement, followed by stars, and galaxies benefiting the least. This suggests that the model's performance is sensitive to data volume for sources with more complex or less abundant features. Although our method has demonstrated impressive classification performance, there is still room for further improvement as spectral data continues to accumulate in the future.

In our extrapolation tests, we divided the data into different magnitude bins based on the $r$-band magnitude and used the data from fainter bins for testing. The results indicate that the model can extrapolate up to 1 mag fainter than the training data. However, this test has inherent limitations in terms of quantitatively analyzing the data and feature space. For example, enforcing equal magnitude intervals for each bin results in varying numbers of training samples across bins. Conversely, ensuring an equal number of samples in each bin alters the magnitude intervals, which in turn affects the structure of the feature space. Additionally, regardless of these considerations, the relative proportions of stars, QSOs, and galaxies unavoidably vary across bins, and controlling the distribution of color features remains challenging.

More importantly, the high-dimensional feature space utilized by the DNN may rely on latent features that are beyond our explicit control or understanding. Despite these challenges, as shown in Table \ref{tab2}, the performance differences across subsets—whether caused by variations in magnitude intervals or sample sizes—are generally within 0.5\% to 4\%. Among these, the 17.5–19.5 and 19–20 magnitude bins are particularly notable because they share identical magnitude intervals and have nearly equal sample sizes (43,905 and 42,849, respectively). The performance differences between these two bins closely align with the expected test variations, suggesting that the extrapolated results remain acceptable within these limits.

While there may be other feature restrictions besides brightness, the testing results from the Gaia and GAMA surveys suggest that their impact is not substantial. These two datasets, apart from the targets being relatively bright ($r <$ 20 mag), may have different feature distributions compared to SDSS, but our model still performs very well on them, indicating that within the permissible brightness range, the model can be well generalized to new data.

Furthermore, some intrinsic issues with KiDS data could also affect the accuracy of the final predictions. For instance, the presence of obvious companions or saturation around some targets (as shown in Figure~\ref{fig11}) may impact the measurement of target flux or the extraction of morphological features, potentially leading to misclassification. Systematically addressing these issues requires comprehensive cleaning and preprocessing of predictive data, which is beyond the scope of this study.

\section{Conclusion} \label{sec:6}

In this work, we have developed an MNN architecture that effectively combines morphological information from images and multi-band photometry for the classification of stars, quasars, and galaxies in the KiDS DR5 survey. By training on SDSS spectroscopic labels and testing on both internal and external datasets, we demonstrate the model's high accuracy, robustness, and ability to generalize to new data. Our key findings are:

\begin{enumerate}
\item On the internal SDSS testing dataset, the model achieves an overall accuracy of 98.76\%, with per-class F1 scores of 98.63\%, 95.61\%, and 99.26\% for stars, QSOs, and galaxies respectively. Raising the output probability threshold can boost purity at the cost of completeness.

\item The model correctly classifies 99.74\% of a pure Gaia star sample, higher than the performance on the SDSS testing dataset, indicating potential mislabeling in SDSS that the network helps correct. It also correctly classifies 99.74\% of GAMA galaxies after accounting for low-redshift contaminants.

    \item We apply the model to classify over 27 million KiDS DR5 sources down to $r=23$ mag. Over 85\% of the resulting catalog has a maximum probability $\geqslant$ 0.9, indicating high classification confidence.

    \item The model's strong performance stems from its ability to extract both morphological and SED features. Normalizing images has little effect, showing the network relies on physical morphological information rather than just flux correction. Compared to an SED-only MLP, the multimodal network achieves significantly better performance.

\end{enumerate}

This work presents a novel and effective approach for classifying astronomical sources by leveraging deep learning to combine imaging and photometric data. The resulting catalog will enable a wide range of galaxy evolution, large-scale structure, and cosmology studies with KiDS DR5. Future work will extend the technique to a wider range of surveys and wavelengths, and explore its potential for identifying unusual objects. Integrating multimodal information with deep learning offers a promising tool for maximizing the scientific return from the next generation of large-scale astronomical surveys.

\vspace{5mm}

We thank the anonymous referee for their valuable comments and suggestions, which greatly improved the quality of this work. We also thank Yuxuan Pang for his discussions. This work is supported by National Key R\&D Program of China (No. 2021YFA1600404), the National Natural Science Foundation of China (NSFC grants No. 12203096, 12303022, 12203050, 12373018, 11991051, and 12203041), Yunnan Fundamental Research Projects (grants NO. 202301AT070358 and 202301AT070339), Yunnan Postdoctoral Research Foundation Funding Project, Special Research Assistant Funding Project of Chinese Academy of Sciences, and the science research grants from the China Manned Space Project (No. CMS-CSST-2021-A06, CMS-CSST-2025-A02, CMS-CSST-2025-A07), Hunan Outstanding Youth Science Foundation(2024JJ2040), NRN acknowledges financial support from the NSFC, Research Fund for Excellent International Scholars (grant No. 12150710511), and from the research grant from China Manned Space Project No. CMS-CSST-2021-A01.

This research made use of the cross-match service provided by CDS, Strasbourg. Based on data products from observations made with ESO Telescopes at the La Silla Paranal Observatory under programme IDs 177.A-3016, 177.A-3017 and 177.A-3018, and on data products produced by Target/OmegaCEN, INAF-OACN, INAF-OAPD and the KiDS production team, on behalf of the KiDS consortium. OmegaCEN and the KiDS production team acknowledge support by NOVA and NWO-M grants. Members of INAF-OAPD and INAF-OACN also acknowledge the support from the Department of Physics \& Astronomy of the University of Padova, and of the Department of Physics of Univ. Federico II (Naples).
Funding for the Sloan Digital Sky Survey IV has been provided by the Alfred P. Sloan Foundation, the U.S. Department of Energy Office of Science, and the Participating Institutions. SDSS-IV acknowledges support and resources from the Center for High Performance Computing  at the University of Utah. The SDSS website is \url{www.sdss4.org}. SDSS-IV is managed by the Astrophysical Research Consortium for the Participating Institutions of the SDSS Collaboration including the Brazilian Participation Group, the Carnegie Institution for Science, Carnegie Mellon University, Center for Astrophysics | Harvard \& Smithsonian, the Chilean Participation Group, the French Participation Group, Instituto de Astrof\'isica de Canarias, The Johns Hopkins University, Kavli Institute for the Physics and Mathematics of the Universe (IPMU) / University of Tokyo, the Korean Participation Group, Lawrence Berkeley National Laboratory, Leibniz Institut f\"ur Astrophysik Potsdam (AIP),  Max-Planck-Institut f\"ur Astronomie (MPIA Heidelberg), Max-Planck-Institut f\"ur Astrophysik (MPA Garching), Max-Planck-Institut f\"ur Extraterrestrische Physik (MPE), National Astronomical Observatories of China, New Mexico State University, New York University, University of Notre Dame, Observat\'ario Nacional / MCTI, The Ohio State University, Pennsylvania State University, Shanghai Astronomical Observatory, United Kingdom Participation Group, Universidad Nacional Aut\'onoma de M\'exico, University of Arizona, University of Colorado Boulder, University of Oxford, University of Portsmouth, University of Utah, University of Virginia, University of Washington, University of Wisconsin, Vanderbilt University, and Yale University. 
This work has made use of data from the European Space Agency (ESA) mission {\it Gaia} (\url{https://www.cosmos.esa.int/gaia}), processed by the {\it Gaia} Data Processing and Analysis Consortium (DPAC, \url{https://www.cosmos.esa.int/web/gaia/dpac/consortium}). Funding for the DPAC has been provided by national institutions, in particular the institutions participating in the {\it Gaia} Multilateral Agreement. GAMA is a joint European-Australasian project based around a spectroscopic campaign using the Anglo-Australian Telescope. 
The GAMA input catalogue is based on data taken from the Sloan Digital Sky Survey and the UKIRT Infrared Deep Sky Survey. Complementary imaging of the GAMA regions is being obtained by a number of independent survey programmes including GALEX MIS, VST KiDS, VISTA VIKING, WISE, Herschel-ATLAS, GMRT and ASKAP providing UV to radio coverage. GAMA is funded by the STFC (UK), the ARC (Australia), the AAO, and the participating institutions. The GAMA website is \url{https://www.gama-survey.org/}.

\vspace{5mm}

\begin{figure*}[!ht]
\centering
\includegraphics[scale=0.45]{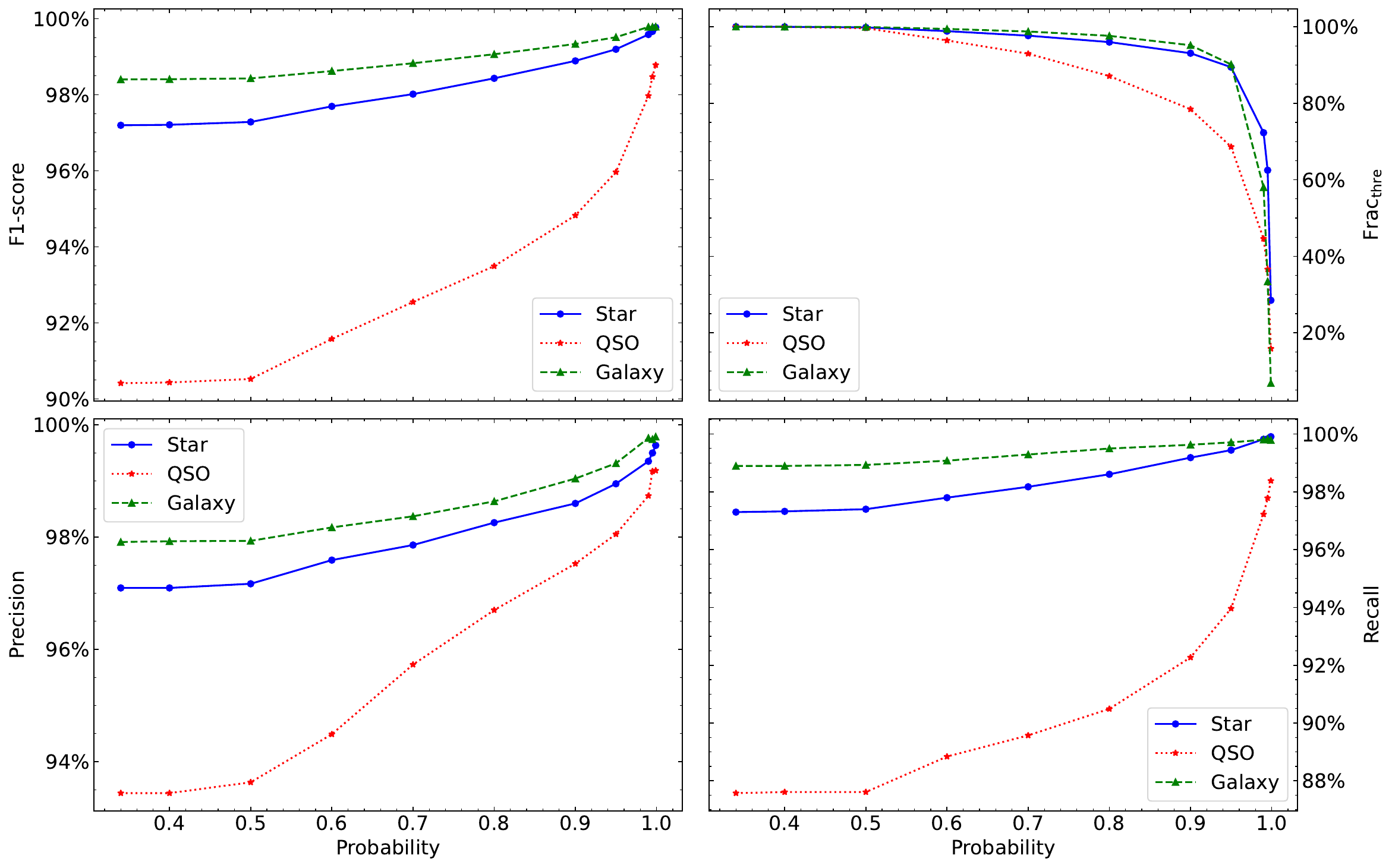}
\caption{The same as Figure~\ref{fig3}, but showing evaluation metrics for classification using only the MLP branch.} 
\label{fig12}
\end{figure*}

\appendix
\section{Performance of MLP} \label{sec:appendixA}
To assess the contribution of morphological information in our MNN, we conducted an independent performance evaluation of the MLP branch. In this test, the MLP branch used only the SED features from the KiDS dataset, which include photometry from 9-bands and 8 adjacent band colors, without integrating the morphological features processed by the CNN branch. The MLP was trained and evaluated using the same training and testing datasets as the full multimodal network. Figure~\ref{fig12} shows the evaluation metrics for the MLP at different output probability thresholds, analogous to Figure~\ref{fig3} for the full multimodal network.

The MLP achieves an overall accuracy of 97.15\% on the testing dataset, with F1 scores of 97.20\%, 90.41\%, and 98.40\% for stars, quasars, and galaxies at the default threshold (Table \ref{tab1}). While still good, these metrics are noticeably lower than the performance of the multimodal network, especially for quasars. The MLP maintains high purity for galaxies and stars as the threshold is raised, but suffers a faster drop in completeness compared to the multimodal network. For quasars, the MLP fails to achieve both high purity and high completeness at any threshold, whereas the multimodal network can reach 99\% purity with 83\% completeness at $p \geqslant 0.99$.

These results demonstrate that the morphological information extracted by the CNN branch contributes significantly to the high performance of the multimodal network, especially when dealing with objects that are difficult to distinguish morphologically. In this way, the MNN effectively leverages complementary information from different data sources to achieve high-accuracy classification of stars, quasars, and galaxies.

\section{Impact of Training Data Size on DNN Performance} \label{sec:appendixB}
DNNs are widely recognized for their ability to model complex, non-linear relationships in data, but they often require large training datasets to achieve optimal performance. To investigate how the size of the training dataset affects the classification performance of our MNN, we conducted a series of experiments. For each experiment, we randomly selected subsets of the original training data with sizes ranging from 5,000 to 100,000 samples. The same network architecture and hyperparameters, as described in Section~\ref{sec:3}, were used for all training runs. Each trained model was then evaluated on the same testing dataset detailed in Section~\ref{sec:4.1.1} to ensure consistency.

\begin{figure}[!ht]
\centering
\includegraphics[scale=0.5]{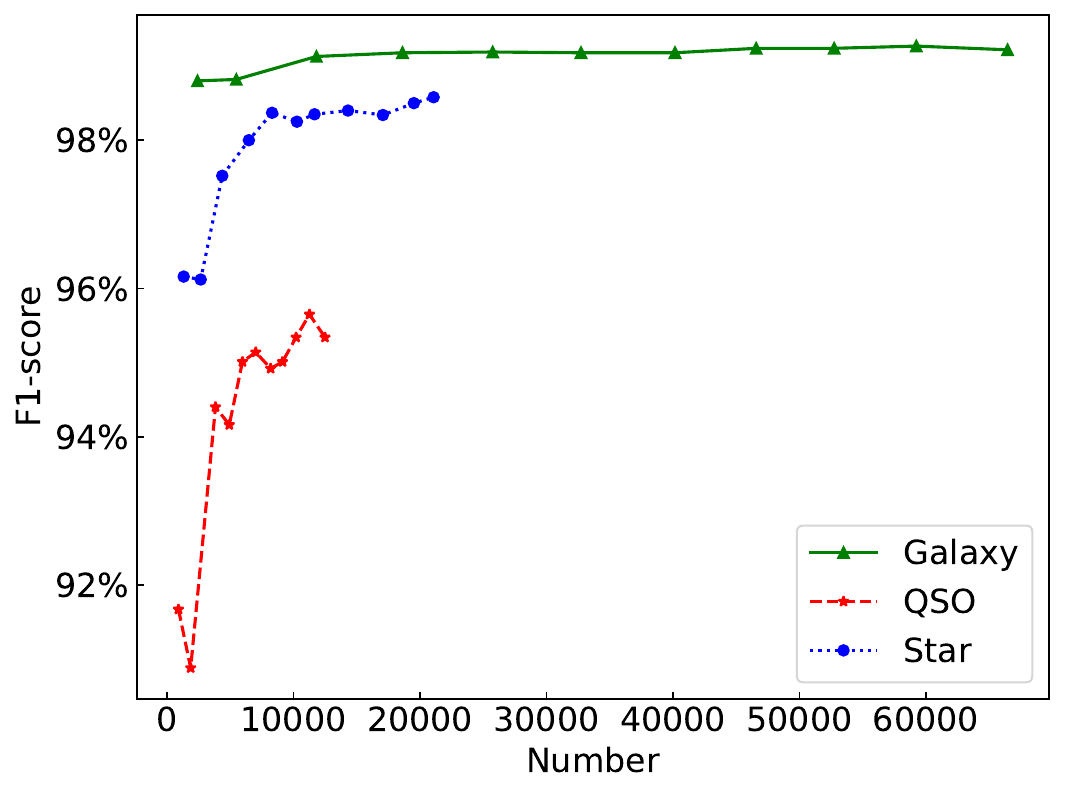}
\caption{F1-scores for stars, quasars, and galaxies as a function of training dataset size. The total training dataset size ranges from 5,000 to 100,000. For each training run, the same validation and testing datasets are used to ensure consistency during the training and evaluation processes.} 
\label{fig13}
\end{figure}

Figure~\ref{fig13} presents the F1-scores for stars, quasars, and galaxies as a function of training data size. The results reveal a clear trend: the performance of the model improves consistently with increasing training data volume for all three source types.

Interestingly, galaxies achieve the highest F1-scores across all training data sizes, even when trained with relatively small datasets. This robustness likely arises from their distinct morphological features, which make them easier to classify. The F1-scores for stars improve steadily with increasing training data size, though the rate of improvement is less pronounced compared to quasars. This suggests that while a moderate amount of data is sufficient to achieve reasonable performance for stars, additional data still provides incremental benefits. Quasars, on the other hand, exhibit the most significant improvement in F1-score as the training dataset grows. This trend indicates that quasars possess more complex or diverse features compared to stars and galaxies, necessitating larger datasets for the network to generalize effectively.

These results underscore the critical role of data volume in training DNNs, particularly for classes with more complex or less distinct features, such as quasars. The performance saturation observed for galaxies, and to a lesser extent for stars, suggests that their feature spaces are relatively simpler for the network to learn, even with smaller training samples. In contrast, the significant improvement in quasar classification with increasing data highlights the importance of expanding the training dataset for these objects in future studies to further enhance model performance.

\begin{thebibliography}{}

\bibitem[Abdurro'uf et al.(2022)]{Ab22} Abdurro'uf, Accetta, K., Aerts, C., et al.\ 2022, \apjs, 259, 35. doi:10.3847/1538-4365/ac4414

\bibitem[Aihara et al.(2018)]{Aihara2018} Aihara, H., Arimoto, N., Armstrong, R., et al.\ 2018, \pasj, 70, S4. doi:10.1093/pasj/psx066

\bibitem[Bai et al.(2019)]{Bai2019} Bai, Y., Liu, J., Wang, S., et al.\ 2019, \aj, 157, 9. doi:10.3847/1538-3881/aaf009

\bibitem[Baldwin et al.(1981)]{Baldwin1981} Baldwin, J.~A., Phillips, M.~M., \& Terlevich, R.\ 1981, \pasp, 93, 5. doi:10.1086/130766

\bibitem[Behroozi et al.(2013)]{Behroozi2013} Behroozi, P.~S., Wechsler, R.~H., \& Conroy, C.\ 2013, \apj, 770, 57. doi:10.1088/0004-637X/770/1/57

\bibitem[Bengio et al.(2012)]{Bengio2012} Bengio, Y., Courville, A., \& Vincent, P.\ 2012, arXiv:1206.5538. doi:10.48550/arXiv.1206.5538

\bibitem[Bertin \& Arnouts(1996)]{Bertin1996} Bertin, E. \& Arnouts, S.\ 1996, \aaps, 117, 393. doi:10.1051/aas:1996164

\bibitem[Bolton et al.(2012)]{Bolton2012} Bolton, A.~S., Schlegel, D.~J., Aubourg, {\'E}., et al.\ 2012, \aj, 144, 144. doi:10.1088/0004-6256/144/5/144

\bibitem[Bond et al.(1996)]{Bond1996} Bond, J.~R., Kofman, L., \& Pogosyan, D.\ 1996, \nat, 380, 603. doi:10.1038/380603a0

\bibitem[Buchs et al.(2019)]{Buchs2019} Buchs, R., Davis, C., Gruen, D., et al.\ 2019, \mnras, 489, 820. doi:10.1093/mnras/stz2162

\bibitem[Breiman(2001)]{Breiman2001} Breiman, L., 2001, Mach. Learn., 45, 5

\bibitem[Cabayol et al.(2019)]{Cabayol2019} Cabayol, L., Sevilla-Noarbe, I., Fern{\'a}ndez, E., et al.\ 2019, \mnras, 483, 529. doi:10.1093/mnras/sty3129

\bibitem[Capaccioli \& Schipani(2011)]{Ca11} Capaccioli, M. \& Schipani, P.\ 2011, The Messenger, 146, 2

\bibitem[Capaccioli et al.(2012)]{Ca12} Capaccioli, M., Schipani, P., de Paris, G., et al.\ 2012, Science from the Next Generation Imaging and Spectroscopic Surveys, 1

\bibitem[Chaini et al.(2023)]{Chaini2023} Chaini, S., Bagul, A., Deshpande, A., et al.\ 2023, \mnras, 518, 3123. doi:10.1093/mnras/stac3336

\bibitem[Chen \& Guestrin (2016)]{Chen2016} Chen, T., \& Guestrin, C. 2016, in Proc. of the 22nd ACM SIGKDD Int. Conf.
on Knowledge Discovery and Data Mining (New York: Association for
Computing Machinery), 785

\bibitem[Chen(2021)]{Chen2021} Chen, Y.~C.\ 2021, \apjs, 256, 34. doi:10.3847/1538-4365/ac13aa

\bibitem[Conroy(2013)]{Conroy2013} Conroy, C.\ 2013, \araa, 51, 393. doi:10.1146/annurev-astro-082812-141017

\bibitem[Cui et al.(2012)]{Cui2012} Cui, X.-Q., Zhao, Y.-H., Chu, Y.-Q., et al.\ 2012, Research in Astronomy and Astrophysics, 12, 1197. doi:10.1088/1674-4527/12/9/003

\bibitem[de Jong et al.(2015)]{de15} de Jong, J.~T.~A., Verdoes Kleijn, G.~A., Boxhoorn, D.~R., et al.\ 2015, \aap, 582, A62. doi:10.1051/0004-6361/201526601

\bibitem[The Dark Energy Survey Collaboration(2005)]{DESCollaboration2005} The Dark Energy Survey Collaboration\ 2005, astro-ph/0510346. doi:10.48550/arXiv.astro-ph/0510346

\bibitem[DESI Collaboration et al.(2016)]{DESICollaboration2016} DESI Collaboration, Aghamousa, A., Aguilar, J., et al.\ 2016, arXiv:1611.00036. doi:10.48550/arXiv.1611.00036

\bibitem[Dieleman et al.(2015)]{Dieleman2015} Dieleman, S., Willett, K.~W., \& Dambre, J.\ 2015, \mnras, 450, 1441. doi:10.1093/mnras/stv632

\bibitem[Driver et al.(2011)]{Dr11} Driver, S.~P., Hill, D.~T., Kelvin, L.~S., et al.\ 2011, \mnras, 413, 971. doi:10.1111/j.1365-2966.2010.18188.x

\bibitem[Driver et al.(2022)]{Dr22} Driver, S.~P., Bellstedt, S., Robotham, A.~S.~G., et al.\ 2022, \mnras, 513, 439. doi:10.1093/mnras/stac472

\bibitem[Dubois et al.(2022)]{Dubois2022} Dubois, J., Fraix-Burnet, D., Moultaka, J., et al.\ 2022, \aap, 663, A21. doi:10.1051/0004-6361/202141729

\bibitem[Edge et al.(2013)]{Ed13} Edge, A., Sutherland, W., Kuijken, K., et al.\ 2013, The Messenger, 154, 32

\bibitem[Elvis et al.(1994)]{Elvis1994} Elvis, M., Wilkes, B.~J., McDowell, J.~C., et al.\ 1994, \apjs, 95, 1. doi:10.1086/192093

\bibitem[Fadely et al.(2012)]{Fadely2012} Fadely, R., Hogg, D.~W., \& Willman, B.\ 2012, \apj, 760, 15. doi:10.1088/0004-637X/760/1/15

\bibitem[Feng et al.(2021)]{Feng2021} Feng, H.-C., Hu, C., Li, S.-S., et al.\ 2021, \apj, 909, 18. doi:10.3847/1538-4357/abd851

\bibitem[Fotopoulou et al.(2016)]{Fotopoulou2016} Fotopoulou, S., Pacaud, F., Paltani, S., et al.\ 2016, \aap, 592, A5. doi:10.1051/0004-6361/201527402

\bibitem[Gaia Collaboration et al.(2016)]{Ga16} Gaia Collaboration, Prusti, T., de Bruijne, J.~H.~J., et al.\ 2016, \aap, 595, A1. doi:10.1051/0004-6361/201629272

\bibitem[Gaia Collaboration et al.(2021)]{Ga21} Gaia Collaboration, Brown, A.~G.~A., Vallenari, A., et al.\ 2021, \aap, 649, A1. doi:10.1051/0004-6361/202039657

\bibitem[Gaia Collaboration et al.(2022)]{Ga22} Gaia Collaboration, Vallenari, A., Brown, A.~G.~A., et al.\ 2022, arXiv:2208.00211

\bibitem[Glorot et al.(2011)]{Glorot2011} 
Glorot, X., Bordes, A., \& Bengio, Y.\ 2011, in *Proceedings of the Fourteenth International Conference on Artificial Intelligence and Statistics*, ed. G.~Gordon, D.~Dunson, \& M.~Dudík, Vol. 15, Proceedings of Machine Learning Research (Fort Lauderdale, FL, USA: PMLR), 315--323.

\bibitem[Gong et al.(2019)]{Gong2019} Gong, Y., Liu, X., Cao, Y., et al.\ 2019, \apj, 883, 203. doi:10.3847/1538-4357/ab391e

\bibitem[Goodfellow et al.(2016)]{Goodfellow2016} 
Goodfellow, I., Bengio, Y., \& Courville, A.\ 2016, Deep Learning (Cambridge, MA: MIT Press)

\bibitem[Guo et al.(2022)]{Guo2022} Guo, Z., Wu, J.~F., \& Sharon, C.~E.\ 2022, arXiv:2212.07881. doi:10.48550/arXiv.2212.07881

\bibitem[Han et al.(2022)]{Han2022} Han, Z., Yuan, X., Li, Z., et al.\ 2022, \procspie, 12188, 1218842. doi:10.1117/12.2637071

\bibitem[He et al.(2015)]{He15} He, K., Zhang, X., Ren, S., et al.\ 2015, arXiv:1512.03385. doi:10.48550/arXiv.1512.03385

\bibitem[Helmi(2020)]{Helmi2020} Helmi, A.\ 2020, \araa, 58, 205. doi:10.1146/annurev-astro-032620-021917

\bibitem[Hubble(1926)]{Hubble1926} Hubble, E.~P.\ 1926, \apj, 64, 321. doi:10.1086/143018

\bibitem[Ilbert et al.(2009)]{Ilbert2009} Ilbert, O., Capak, P., Salvato, M., et al.\ 2009, \apj, 690, 1236. doi:10.1088/0004-637X/690/2/1236

\bibitem[Ivezi{\'c} et al.(2019)]{Ivezic2019} Ivezi{\'c}, {\v{Z}}., Kahn, S.~M., Tyson, J.~A., et al.\ 2019, \apj, 873, 111. doi:10.3847/1538-4357/ab042c

\bibitem[Kelvin et al.(2014)]{Kelvin2014} Kelvin, L.~S., Driver, S.~P., Robotham, A.~S.~G., et al.\ 2014, \mnras, 439, 1245. doi:10.1093/mnras/stt2391

\bibitem[Khramtsov et al.(2019)]{Khramtsov2019} Khramtsov, V., Sergeyev, A., Spiniello, C., et al.\ 2019, \aap, 632, A56. doi:10.1051/0004-6361/201936006

\bibitem[Khramtsov et al.(2021)]{Khramtsov2021} Khramtsov, V., Spiniello, C., Agnello, A., et al.\ 2021, \aap, 651, A69. doi:10.1051/0004-6361/202040131

\bibitem[Kormendy \& Ho(2013)]{Kormendy2013} Kormendy, J. \& Ho, L.~C.\ 2013, \araa, 51, 511. doi:10.1146/annurev-astro-082708-101811

\bibitem[Kuijken et al.(2019)]{Ku19} Kuijken, K., Heymans, C., Dvornik, A., et al.\ 2019, \aap, 625, A2. doi:10.1051/0004-6361/201834918

\bibitem[Kuijken et al.(2015)]{Ku15} Kuijken, K., Heymans, C., Hildebrandt, H., et al.\ 2015, \mnras, 454, 3500. doi:10.1093/mnras/stv2140

\bibitem[Laureijs et al.(2011)]{Laureijs2011} Laureijs, R., Amiaux, J., Arduini, S., et al.\ 2011, arXiv:1110.3193. doi:10.48550/arXiv.1110.3193

\bibitem[LeCun et al.(2015)]{Lecun2015} LeCun, Y., Bengio, Y., \& Hinton, G.\ 2015, \nat, 521, 436. doi:10.1038/nature14539

\bibitem[Lecun et al.(1998)]{Lecun1998} Lecun, Y., Bottou, L., Bengio, Y., \& Haffner, P. 1998, Proc. IEEE, 86, 2278. doi: 10.1109/5.726791.

\bibitem[Li et al.(2019)]{Li2019} Li, R., Shu, Y., Su, J., et al.\ 2019, \mnras, 482, 313. doi:10.1093/mnras/sty2708

\bibitem[Li et al.(2022)]{Li2022} Li, R., Napolitano, N.~R., Feng, H., et al.\ 2022, \aap, 666, A85. doi:10.1051/0004-6361/202244081

\bibitem[Liu et al.(2022)]{Liu2022} Liu, C., Gebhardt, K., Cooper, E.~M., et al.\ 2022, \apj, 935, 2, 132. doi:10.3847/1538-4357/ac8054

\bibitem[Logan \& Fotopoulou(2020)]{Logan2020} Logan, C.~H.~A. \& Fotopoulou, S.\ 2020, \aap, 633, A154. doi:10.1051/0004-6361/201936648

\bibitem[L{\'o}pez-Sanjuan et al.(2019)]{Lopez-Sanjuan2019} L{\'o}pez-Sanjuan, C., V{\'a}zquez Rami{\'o}, H., Varela, J., et al.\ 2019, \aap, 622, A177. doi:10.1051/0004-6361/201732480

\bibitem[Lyke et al.(2020)]{Lyke2020} Lyke, B.~W., Higley, A.~N., McLane, J.~N., et al.\ 2020, \apjs, 250, 8. doi:10.3847/1538-4365/aba623

\bibitem[MacGillivray et al.(1976)]{MacGillivray1976} MacGillivray, H.~T., Martin, R., Pratt, N.~M., et al.\ 1976, \mnras, 176, 265. doi:10.1093/mnras/176.2.265

\bibitem[McCulloch et al.(1943)]{McCulloch1943} McCulloch, W.~S., Pitts W., 1943, Bull. Math. Biophys., 5, 115. doi:10.1007/BF02478259

\bibitem[Nair \& Hinton(2010)]{Nair2010} 
Nair, V., \& Hinton, G.~E.\ 2010, in *Proceedings of the 27th International Conference on Machine Learning* (ICML'10) (Haifa, Israel: Omnipress), 807–814. ISBN: 9781605589077. doi:10.5555/3104322.3104425

\bibitem[Nakazono et al.(2021)]{Nakazono2021} Nakazono, L., Mendes de Oliveira, C., Hirata, N.~S.~T., et al.\ 2021, \mnras, 507, 5847. doi:10.1093/mnras/stab1835

\bibitem[Nakoneczny et al.(2019)]{Nakoneczny2019} Nakoneczny, S., Bilicki, M., Solarz, A., et al.\ 2019, \aap, 624, A13. doi:10.1051/0004-6361/201834794

\bibitem[Nakoneczny et al.(2021)]{Nakoneczny2021} Nakoneczny, S.~J., Bilicki, M., Pollo, A., et al.\ 2021, \aap, 649, A81. doi:10.1051/0004-6361/202039684

\bibitem[Palanque-Delabrouille et al.(2016)]{PalanqueDelabrouille2016} Palanque-Delabrouille, N., Magneville, C., Y{\`e}che, C., et al.\ 2016, \aap, 587, A41. doi:10.1051/0004-6361/201527392

\bibitem[Peters et al.(2015)]{Peters2015} Peters, C.~M., Richards, G.~T., Myers, A.~D., et al.\ 2015, \apj, 811, 95. doi:10.1088/0004-637X/811/2/95

\bibitem[Richards et al.(2002)]{Richards2002} Richards, G.~T., Fan, X., Newberg, H.~J., et al.\ 2002, \aj, 123, 2945. doi:10.1086/340187

\bibitem[Salpeter(1964)]{Salpeter1964} Salpeter, E.~E.\ 1964, \apj, 140, 796. doi:10.1086/147973

\bibitem[Salvato et al.(2009)]{Salvato2009} Salvato, M., Hasinger, G., Ilbert, O., et al.\ 2009, \apj, 690, 1250. doi:10.1088/0004-637X/690/2/1250

\bibitem[Salvato et al.(2022)]{Salvato2022} Salvato, M., Wolf, J., Dwelly, T., et al.\ 2022, \aap, 661, A3. doi:10.1051/0004-6361/202141631

\bibitem[Sanders et al.(1989)]{Sanders1989} Sanders, D.~B., Phinney, E.~S., Neugebauer, G., et al.\ 1989, \apj, 347, 29. doi:10.1086/168094

\bibitem[Schmidt(1963)]{Schmidt1963} Schmidt, M.\ 1963, \nat, 197, 1040. doi:10.1038/1971040a0

\bibitem[Scranton et al.(2002)]{Scranton2002} Scranton, R., Johnston, D., Dodelson, S., et al.\ 2002, \apj, 579, 48. doi:10.1086/342786

\bibitem[Sevilla-Noarbe et al.(2018)]{Sevilla-Noarbe2018} Sevilla-Noarbe, I., Hoyle, B., March{\~a}, M.~J., et al.\ 2018, \mnras, 481, 5451. doi:10.1093/mnras/sty2579

\bibitem[Skrutskie et al.(2006)]{Skrutskie2006} Skrutskie, M.~F., Cutri, R.~M., Stiening, R., et al.\ 2006, \aj, 131, 1163. doi:10.1086/498708

\bibitem[Smith \& Geach(2023)]{Smith2023} Smith, M.~J. \& Geach, J.~E.\ 2023, Royal Society Open Science, 10, 221454. doi:10.1098/rsos.221454

\bibitem[Szklen{\'a}r et al.(2020)]{Szklenar2020} Szklen{\'a}r, T., B{\'o}di, A., Tarczay-Neh{\'e}z, D., et al.\ 2020, \apjl, 897, L12. doi:10.3847/2041-8213/ab9ca4

\bibitem[Tadaki et al.(2020)]{Tadaki2020} Tadaki, K.-. ichi ., Iye, M., Fukumoto, H., et al.\ 2020, \mnras, 496, 4276. doi:10.1093/mnras/staa1880

\bibitem[Taylor(2005)]{Ta05} Taylor, M.~B.\ 2005, Astronomical Data Analysis Software and Systems XIV, 347, 29

\bibitem[Verro et al.(2022)]{Verro2022} Verro, K., Trager, S.~C., Peletier, R.~F., et al.\ 2022, \aap, 660, A34. doi:10.1051/0004-6361/202142388

\bibitem[Wang et al.(2023)]{Wang2023} Wang, T., Liu, G., Cai, Z., et al.\ 2023, Science China Physics, Mechanics, and Astronomy, 66, 109512. doi:10.1007/s11433-023-2197-5

\bibitem[Wright et al.(2010)]{Wright2010} Wright, E.~L., Eisenhardt, P.~R.~M., Mainzer, A.~K., et al.\ 2010, \aj, 140, 1868. doi:10.1088/0004-6256/140/6/1868

\bibitem[Xie et al.(2023)]{Xie2023} Xie, L., Napolitano, N.~R., Guo, X., et al.\ 2023, Science China Physics, Mechanics, and Astronomy, 66, 129513. doi:10.1007/s11433-023-2173-8

\bibitem[Yang \& Shen(2023)]{Yang2023} Yang, Q. \& Shen, Y.\ 2023, \apjs, 264, 9. doi:10.3847/1538-4365/ac9ea8

\bibitem[Yee(1991)]{Yee1991} Yee, H.~K.~C.\ 1991, \pasp, 103, 396. doi:10.1086/132834

\bibitem[York et al.(2000)]{Yo00} York, D.~G., Adelman, J., Anderson, J.~E., et al.\ 2000, \aj, 120, 1579. doi:10.1086/301513

\bibitem[Zeraatgari et al.(2024)]{Zeraatgari2024} Zeraatgari, F.~Z., Hafezianzadeh, F., Zhang, Y., et al.\ 2024, \mnras, 527, 4677. doi:10.1093/mnras/stad3436

\bibitem[Zhang et al.(2016)]{Zhang2016} Zhang, C., Bengio, S., Hardt, M., et al.\ 2016, arXiv:1611.03530. doi:10.48550/arXiv.1611.03530

\bibitem[Zhang et al.(2023)]{Zhang2023} Zhang, X., Green, G.~M., \& Rix, H.-W.\ 2023, \mnras, 524, 1855. doi:10.1093/mnras/stad1941

\bibitem[Zhou et al.(2021)]{Zhou2021} Zhou, X., Gong, Y., Meng, X.-M., et al.\ 2021, \apj, 909, 53. doi:10.3847/1538-4357/abda3e

\bibitem[Zuntz et al.(2018)]{Zuntz2018} Zuntz, J., Sheldon, E., Samuroff, S., et al.\ 2018, \mnras, 481, 1149. doi:10.1093/mnras/sty2219

\end{thebibliography}
\end{document}